\documentclass[conference]{IEEEtran}
\IEEEoverridecommandlockouts
\usepackage{cite, comment}
\usepackage{algorithmic}
\usepackage[ruled]{algorithm2e} 
\usepackage{graphicx}
\usepackage{times}
\usepackage{array}
\usepackage{stfloats}
\usepackage{rotating,threeparttable,booktabs}
\usepackage{dcolumn}
\usepackage{multirow}
\usepackage{color}
\usepackage{makecell}
\usepackage{epstopdf}
\usepackage{cuted}
\usepackage{subfigure}
\usepackage{multicol}
\usepackage{amsmath,amssymb}
\usepackage[final]{hyperref}
\hypersetup{
	colorlinks=true,       
	linkcolor=blue,        
	citecolor=blue,        
	filecolor=magenta,     
	urlcolor=blue         
}
\def\BibTeX{{\rm B\kern-.05em{\sc i\kern-.025em b}\kern-.08em
    T\kern-.1667em\lower.7ex\hbox{E}\kern-.125emX}}

\newcommand{\cwu}[1]{{\textcolor{red}{CWu: #1}}}

\newcommand{\mf}[1]{{\textcolor{blue}{MF: #1}}}

\begin{document}

\title{\huge Optimizing Distributed Deployment of Mixture-of-Experts Model Inference in Serverless Computing}

\author{\IEEEauthorblockN{Mengfan Liu\IEEEauthorrefmark{1}, Wei Wang\IEEEauthorrefmark{2}, and Chuan Wu\IEEEauthorrefmark{1}
}
\IEEEauthorblockA{\IEEEauthorrefmark{1}Department of Computer Science,
The University of Hong Kong}
\IEEEauthorblockA{\IEEEauthorrefmark{2}Department of Computer Science and Engineering,
The Hong Kong University of Science and Technology\\
Email: ml621@connect.hku.hk, weiwa@cse.ust.hk, cwu@cs.hku.hk}}

\maketitle

\pdfoutput=1
\begin{abstract}
With the advancement of serverless computing, running machine learning (ML) inference services over a serverless platform has been advocated, given its labor-free scalability and cost effectiveness. 
Mixture-of-Experts (MoE) models have been a dominant type of model architectures to enable large models nowadays, 
with parallel expert networks. Serving large MoE models on serverless computing is potentially beneficial, but has been underexplored due to substantial challenges in handling the skewed expert popularity and scatter-gather communication bottleneck in MoE model execution, for cost-efficient serverless MoE deployment and performance guarantee. 
We study optimized MoE model deployment and distributed inference serving on a serverless platform, that effectively predict expert selection, pipeline communication with model execution, and minimize the overall billed cost of serving MoE models. Especially, we propose a Bayesian optimization framework with multi-dimensional ${\bf \epsilon}$-greedy search to learn expert selections and optimal MoE deployment achieving optimal billed cost, including: 1) a Bayesian decision-making method for predicting expert popularity; 2) flexibly pipelined scatter-gather communication; and 3) an optimal model deployment algorithm for distributed MoE serving.
Extensive experiments on AWS Lambda show that our designs reduce the billed cost of all MoE layers by at least 75.67$\%$ compared to CPU clusters while maintaining satisfactory inference throughput.  
As compared to LambdaML in serverless computing 
, our designs achieves 43.41$\%$ lower cost with a throughput decrease of at most 18.76$\%$. 
\end{abstract}

\section{Introduction}

Serverless computing is a cloud computing paradigm that the cloud provider elastically manages the provisioning of resources (servers, functions, containers, storage, etc.) to deploy services according to their user demand \cite{jin2023ditto}. 
Serverless computing has been used for serving data analytic applications such as web services 
\cite{jin2023ditto}\cite{zhang2021caerus}. 
In recent years, there has been an increasing trend in adopting serverless computing for machine learning (ML) services, particularly for model inference serving. For example, Gillis \cite{yu2021gillis} studies model partitioning and scheduling for large deep neural network (DNN) inference on AWS
Lambda \cite{Amazon} and Google Cloud Functions \cite{Google}; 
TETRIS \cite{li2022tetris} designs a memory-efficient serverless computing runtime 
for DNN inference. 

Deploying ML inference services on a serverless platform is more appealing than using traditional GPU/CPU clusters for several reasons. {\em First}, it frees ML developers from managing hardware resources and virtual machine/container environments, simplifying service deployment and maintenance \cite{mampage2022holistic}. {\em Second}, its pay-as-you-go pricing model ensures cost efficiency by charging only for resources actively used in fine granularity, avoiding unnecessary costs for idle resources \cite{kounev2023serverless}. {\em Third}, 
serverless functions have been provided to support parallelisms needed for large-scale ML model inference like AWS
Lambda Functions 
\cite{Amazon}, Google Cloud Functions \cite{Google}, Azure Functions 
\cite{Azure}, and Alibaba Cloud Functions \cite{Alibaba}. State-of-the-art commercial serverless platforms largely support CPU services \cite{Amazon}\cite{Google}\cite{Azure}.
Though GPU-based model inference has been preferred for high serving performance, using CPUs for inference serving has been a viable alternative, given that high-caliber GPUs are often in shortage, CPUs are more available and provide substantial cost savings, while being able to meet application service level objectives (SLOs) \cite{he2024distributed}. 

The Mixture-of-Experts (MoE) models have been a dominant type of model architectures to enable large models nowadays \cite{fedus2022switch}, 
achieving high model capacity without increasing computation 
\cite{du2022glam}\cite{rajbhandari2022deepspeed}. To build a large model using the MoE architecture, layers in a representative DNN model (e.g., Transformer) are replaced by MoE layers. Each MoE layer includes a gating network and multiple parallel expert networks. 
During model inference, each input token to an MoE layer is first evaluated by the gating network, which determines the most relevant expert(s) to handle the token \cite{fedus2022switch}. Then the token is routed to the selected expert(s) for computation, and the processing results are aggregated to produce the MoE layer output. 
MoE models have been widely used to 
serve various tasks, 
e.g., GLaM \cite{du2022glam} 
for speech recognition 
and SwitchTransformer \cite{fedus2022switch} 
for 
text generation. 

Serving a large MoE model is resource intensive as it requires substantial memory to deploy the parallel experts.
DeepSpeed \cite{rajbhandari2022deepspeed} and Lina \cite{li2023accelerating} adopt expert parallelism to accelerate inference with an expert assigned to a device (mostly GPU) and all-to-all communication to receive inputs from other devices. After deployment, devices incur costs even when idle, making GPU/CPUs more costly than using a serverless platform.

We advocate serving MoE models in a CPU-based serverless platform, for labor and cost-efficient management 
of inference serving. 
Serving large MoE models on serverless computing
has been underexplored. 
The main target for MoE model deployment in a serverless platform is to minimize the billed cost of all MoE layers in serving, for memory usage and execution time of serverless functions that run the MoE layers 
\cite{Amazon}\cite{Google}\cite{Azure}\cite{Alibaba}. 
Two major challenges arise for cost-minimal distributed deployment of MoE model inference in a serverless platform.

{\em First}, serverless functions are typically deployed with memory size configured before the service runs, and the skewed, 
unknown-beforehand expert selection of input tokens 
complicates proactive memory configuration of the functions. In MoE serving, some experts (each run as a serverless function) receive many tokens for processing, 
while others do not. 
Intuitively, popular experts should be run on serverless functions with larger memory while non-popular ones use less memory. 
Existing MoE serving solutions \cite{he2021fastmoe}\cite{zhai2023smartmoe} in GPU/CPU clusters 
decide resource assignment for experts 
during MoE inference, which is infeasible in serverless platforms. Deploying a serverless function takes several minutes. The first time after a serverless function is deployed, it takes a long time for the function to start execution, due to resource initialization 
(i.e., the cold start issue) \cite{zhang2021caerus}. This would cause long delays in MoE serving if its serverless functions are deployed according to the current demand, degrading efficiency and throughput of inference serving. 
Therefore, the key challenge here is to efficiently and accurately estimate expert popularity before the MoE inference service starts in a serverless platform. This enables optimizing memory configurations for serverless functions, 
thereby decreasing the billed cost of MoE serving. 

{\em Second}, the scatter-gather communication for token-to-expert routing and expert processing result aggregation at MoE layers is time-consuming, 
that may block subsequent operations as non-MoE layers must wait for all experts to complete their computation and communication
\cite{rajbhandari2022deepspeed}. 
Existing proposals on redesigning MoE scatter-gather communication with pipelining \cite{shi2023pipemoe}\cite{zhang2023mpipemoe} in CPU/GPU clusters are inadequate in serverless platforms. 
Direct inter-function transfers in a serverless platform are constrained by platform-specified maximum data transfer size (i.e., payload size), while indirect transfers via external storage (e.g., S3 bucket for AWS Lambda \cite{Amazon}) take longer as the data must be saved to external storage and then retrieved. 
Pipelining data transfer with model execution 
is infeasible with direct transfers: a serverless function retains no data between invocations (i.e., stateless property) and direct data transfers from other functions require re-invocation of the function each time; model parameters that a serverless function uses are not retained during direct transfers and hence need to be reloaded for each re-invocation, resulting in significant time and memory waste. 
Indirect transfers between serverless functions rely on external storage, incurring longer communication time and higher cost, adding complexity to pipelining design.
This calls for novel communication designs tailored to MoE inference in serverless platforms.

Tackling these challenges, we design a serverless MoE inference solution that effectively predicts
expert popularity, pipelines MoE communication with model
execution, and optimally deploys MoE model for distributed inference, that minimize the billed cost of all MoE layers in MoE model serving. 
We propose a Bayesian optimization (BO) framework with multiple ${\bf \epsilon}$-greedy search (GS) to learn expert popularity and optimize MoE
deployment for billed cost minimization. 
Our main 
contributions are summarized as follows:


$\triangleright$ We design a novel Bayesian decision-making approach for expert selection prediction, including a comprehensive token feature design, a novel posterior calculation approach, and an adjustable key-value dataset table. We analyze the MoE inference process to extract relevant token features. The posterior calculation incorporates real request distributions to refine the posterior using profiled data. The key-value dataset table is adjusted 
with new key-value pairs according to feedback from model inference, which are used to update the profiled data probabilities and improve prediction accuracy.

$\triangleright$ We propose several scatter-gather communication designs for a serverless platform
: indirect transfers with flexible pipeline operations 
via external storage,  simple indirect transmissions without pipelining, 
and simple direct invocation of serverless functions without pipelining. As different designs perform the best in terms of execution time under different scenarios, 
selecting a proper scatter-gather communication design affects billed cost saving in serverless MoE inference. 

$\triangleright$ We formulate optimal deployment of distributed MoE inference on a serverless platform into a mixed-integer quadratically constrained programming (MIQCP) problem, which chooses one of the proposed scatter-gather communication designs,
sets memory configurations of expert serverless functions,
and determines the number of function replicas. 
We design an optimal deployment selection (ODS) algorithm, achieving a billed cost of all MoE layers upper bounded by a constant ratio of the optimal solutions. 


$\triangleright$ We propose a BO framework with multiple ${\bf \epsilon}$-GS to optimize 
expert selection predictions and distributed deployment of the MoE model. The BO framework iteratively adjusts the key-value dataset table for expert selection prediction using 
feedback billed cost of all MoE layers in serving. 
The expert predictions then serve as input for optimal MoE model deployment, which further provides cost feedback for dataset table update. The multiple ${\bf \epsilon}$-GS balances exploration of different key-value pairs in the dataset table and exploitation of 
high-performing key-value pairs, in dataset table update. 

$\triangleright$ We implement our designs on AWS Lambda using Pytorch and Optuna packages. 
Extensive experiments 
show that our designs reduce the billed cost of all MoE layers by at least 75.67$\%$ compared to inference over CPU clusters, while maintaining a satisfactory inference throughput. 
by at least 43.41$\%$ compared to LambdaML \cite{lambdaml} in serverless computing, with a throughput decrease of at most 18.76$\%$.

\section{Background and Motivation}
\subsection{Serverless Computing}\label{secIIA}

\begin{figure}[htbp]
\centerline{\includegraphics[width=0.5\textwidth]{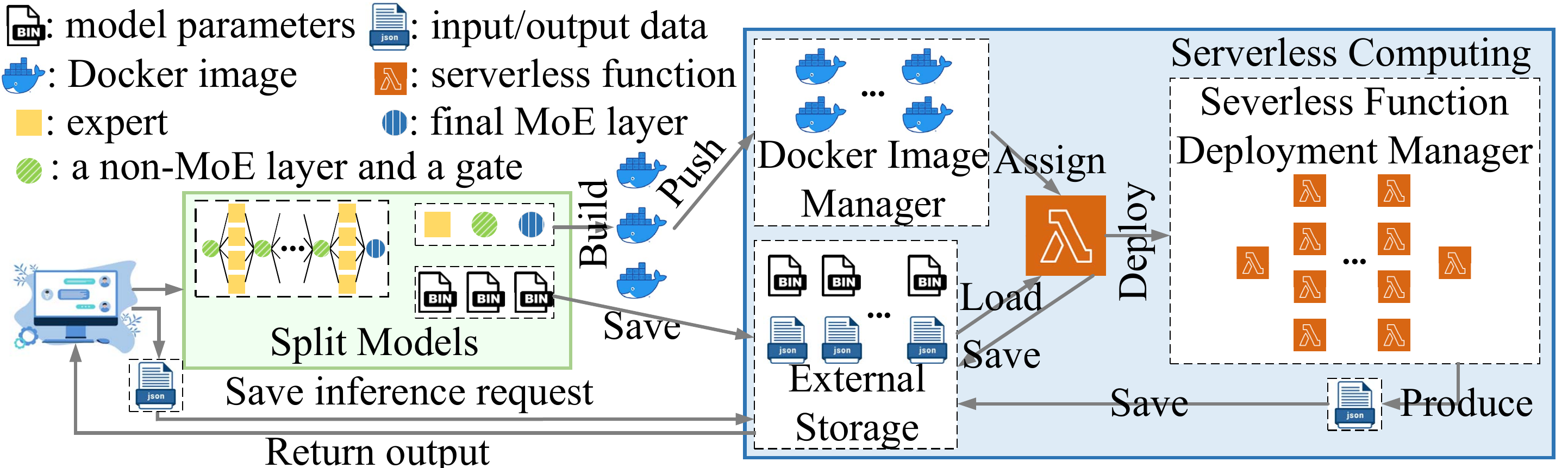}}
\vspace{-3mm}
\caption{Overview of MoE model deployment on a serverless platform.}
\label{ServerlessComputing}
\vspace{-4mm}
\end{figure}

Serverless computing is a cloud computing paradigm that provides Functions as a Service (FaaS). A serverless function is a piece of code running 
in a cloud infrastructure for applications or services, without the need of the developer to provision or manage the resources. These functions can be triggered by various events \cite{kounev2023serverless}, such as inference requests. 
Serverless functions are stateless, that they retain no data from their previous execution. 
Serverless functions obtain data inputs mainly in two ways: from output 
of other functions or from an external storage 
A function can directly transfer output to another function as input when invoking the latter, and a {\em payload} size limits the maximal
data transfer size between functions.  
When the data exceeds the payload size, external storage is used for relaying data between two functions \cite{jin2023ditto}\cite{yu2021gillis}. 
External storage can also store other data needed by serverless functions, such as 
model parameters.

Serverless functions, along with external storage, Docker image manager (e.g., ECR \cite{ecr} for Amazon Lambda \cite{Amazon}), and serverless function deployment manager (e.g., step function \cite{step_func} on Amazon Lambda \cite{Amazon}) 
can be used for ML inference. 
As shown in Fig.~\ref{ServerlessComputing}, deployment of an MoE model for inference serving on a serverless platform involves several steps. First, the MoE model is partitioned: the MoE layer adopts expert parallelism, while the non-MoE layers are grouped and distributed according to model parallelism. 
Next, each model partition is built as a Docker image, which is then pushed to the Docker image manager, with its model parameters stored in external storage. Finally, each Docker image is assigned to a serverless function and these functions are deployed into the serverless platform by a serverless function manager. After deployment, inference requests from service users are stored in external storage and retrieved by the deployed MoE model for inference serving. During inference, each serverless function loads its model parameters and intermediate computation results from external storage, and saves intermediate results back to external storage. 

Current commercial serverless computing platforms (e.g., AWS Lambda \cite{Amazon}, Google Cloud Functions \cite{Google}, Azure \cite{Azure}, and Alibaba Cloud Functions \cite{Alibaba}) are CPU-based. 
Therefore, we focus on CPU-based serverless platforms in this paper.


\vspace{1mm}
\noindent\textbf{Function 
configuration.} Memory is the principal lever available to developers for controlling the performance of a serverless function. The memory configuration of a serverless function determines its computing speed, as more memory corresponds to more virtual CPUs 
\cite{Amazon}\cite{Google}. 
Users can configure the amount of memory to a serverless function and the serverless platform automatically allocates resources based on this configuration during function deployment. 
Commercial serverless computing providers \cite{Amazon}\cite{Google}\cite{Azure}\cite{Alibaba}
 generally charge users on the used memory during 
running time of a serverless function at the unit of GB-second (i.e., GBs). External storage and Docker image manager is charged on the size of stored objects, and severless function deployment manager usage is charged on the number of function invocations.
We focus on the billed cost of serverless functions as it represents 
the primary 
cost 
for MoE model inference, 
given a fixed model size and inference request workload. 

\subsection{MoE Inference} 
The gating network routes tokens to experts based on token features, 
e.g., position in the inference request sequence, meaning of a word token, its roles in the sequence (subject, object, pronoun, verb, etc.) \cite{li2023accelerating}.  
This routing generally does not restrict the number of tokens to be processed by each expert, resulting in skewed popularity and unbalanced workloads among experts. GShard \cite{lepikhin2020gshard} enforces a threshold for the maximal number of tokens processed by one expert.
Zhou et al.~\cite{zhou2022mixture} 
allow each expert to select the top-$k$ tokens and 
operate with a fixed bucket size.
Pre-gated MoE \cite{hwang2023pre} modifies the role of a gate function to preemptively 
select the experts to be activated for the next MoE block. 
These proposals intervene in token-to-expert selection 
and hence may change model results \cite{lepikhin2020gshard, zhou2022mixture, hwang2023pre}. 
We do not modify token-to-expert routing decisions in our 
MoE inference serving.

\vspace{1mm}
\noindent\textbf{Distributed MoE inference.} 
Expert parallelism is commonly adopted that allocates one device 
for each expert \cite{rajbhandari2022deepspeed}\cite{li2023accelerating}. 
For data parallelism in distributed MoE serving, each request batch is distributed across multiple devices 
for simultaneous processing \cite{fedus2022switch}\cite{lepikhin2020gshard}. 
These parallelisms necessitate 
communication across multiple devices for input distribution and output aggregation. 
FasterMoE \cite{he2022fastermoe} applies tensor slicing and pipelining design to overlap all-to-all communication and computation in MoE layers. PipeMoE \cite{shi2023pipemoe} studies the pipeline degree for tensor slicing to adaptively pipeline communication and computation. 
MpipeMoE \cite{zhang2023mpipemoe} extends tensor slicing to multiple dimensions of the data, further expediting inference. The designs rely on the hardware architecture of GPU/CPU clusters, and thus cannot be adopted on serverless platforms. 

\vspace{1mm}
\noindent\textbf{ML services in serverless platforms.} 
Extensive research has focused on optimizing ML model serving in serverless platforms. Amps-inf \cite{jarachanthan2021amps} studies model partitioning to minimize inference costs. 
Ali et al.~\cite{ali2020batch} design a request batch queue to reduce costs, while INFless \cite{yang2022infless} reduces inference latency 
by optimizing batch queuing, execution time, and cold start rates. 
Gillis \cite{yu2021gillis} adopts model partitioning and scheduling for low-latency large DNN model inference, while ServerlessLLM \cite{fu2024serverlessllm} focuses on fast multi-tier checkpoint loading and optimized startup-time scheduling. 
DNN serving on serverless platforms has been proven effective in meeting serving SLOs and 
handling 
varying workloads.
However, there is a lack of designs for efficient MoE inference on a serverless platform. 

\subsection{Opportunities and challenges}
\begin{figure}[!t]
\centerline{\includegraphics[width=0.48\textwidth]{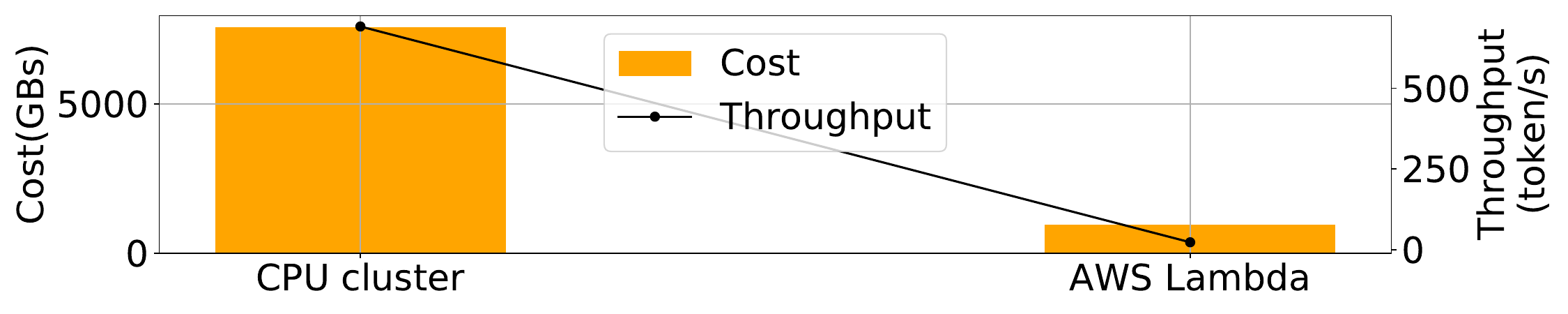}}
\vspace{-3mm}
\caption{Billed cost of all MoE layers and inference throughput of a GPT-2-based MoE model. 
}
\label{fig_cpu_throughput}
\vspace{-5mm}
\end{figure}
\noindent\textbf{Opportunity: MoE model inference in a serverless platform can bring substantial cost reduction and satisfactory inference performance}, as compared to CPU cluster-based inference serving. 
On a serverless platform,  
expert networks can be individually assigned to serverless functions with different memory size configurations, so that more resources are rent for workload-heavy experts and less for workload-light experts. 
Further, inference functions are invoked on-demand at fine-grained timescales (i.e., milliseconds) \cite{Amazon}\cite{Google}.
Serverless functions 
are only billed on assigned memory when executing; no cost for idle resources is incurred. However, the cost of a CPU cluster typically depends on the amount of resources over a fixed coarse-grained period (e.g., per month or per hour), so that costs may still be incurred for idle resources.
Fig.~\ref{fig_cpu_throughput} shows the billed cost of all MoE layers and inference throughput, when a GPT-2-based MoE model serves 10,240 tokens from the Enwiki8 \cite{enwik8} dataset. 
The CPU cluster uses two 64-core AMD EPYC CPUs with 512GB of DRAM for the entire MoE model. 
Each serverless function for MoE inference on AWS Lambda is allocated 3008 MB of memory. The billed cost of all MoE layers on AWS Lambda is significantly lower than on the CPU cluster. The throughput of MoE inference on the serverless platform is 22.9 tokens per second, substantially exceeding the human reading speed of 3.3 tokens per second \cite{kotani2011machine}. 
Hence, serverless-based MoE inference presents a viable solution.
Two challenges exist on enabling efficient deployment and serving of an MoE model on a serverless platform.

\begin{figure}[!t]
\centerline{\includegraphics[width=0.45\textwidth]{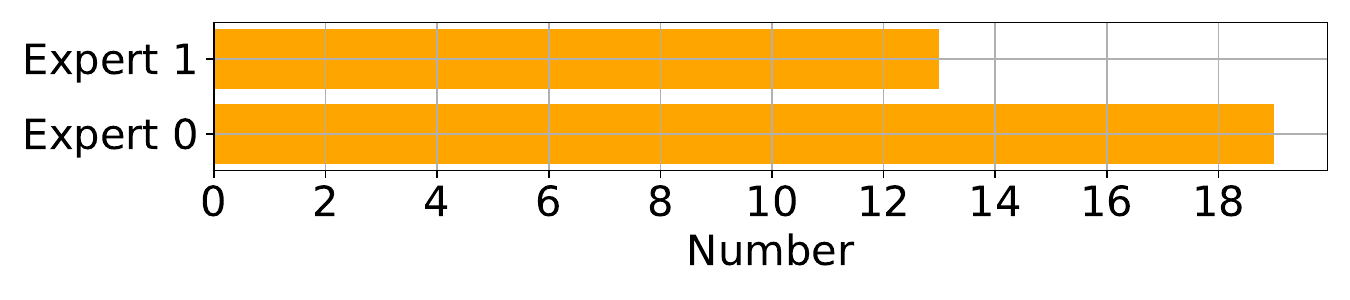}}
\vspace{-3mm}
\caption{Number of tokens with token ID 10424 (from the Enwiki8 dataset) routed to different experts at the 2nd MoE layer in Bert-based MoE model.
}
\label{fig_tid}
\vspace{-5mm}
\end{figure}

\vspace{1mm}
\noindent\textbf{Challenge 1: Skewed, unknown-beforehand expert popularity prevents proactive, accurate memory configuration of serverless functions.} Expert selection is 
unknown before actual token processing by the gating network. 
Dynamic resource allocation, 
which decides computational resources for each expert based on real-time expert popularity during inference, has been advocated for MoE serving in GPU/CPU clusters \cite{he2021fastmoe}\cite{he2022fastermoe}. 
Serverless functions require a long time to deploy (e.g., 1 minute or longer) and to start execution (e.g., 5 seconds or longer) 
after deployed \cite{zhang2021caerus}; 
on-demand dynamic function redeployment is largely infeasible, which would substantially slow down inference serving. This necessitates 
proper configuration of memory sizes when deploying serverless functions before service starts, requiring prior knowledge of expert popularity. Improper memory configuration may either fail to meet the memory demand during inference, or 
incur higher billed costs due to unnecessary memory usage.
FlexMoE~\cite{nie2023flexmoe} and Prophet~\cite{wang2023prophet} use average historical expert popularity 
to adjust resources among experts.
Lina~\cite{li2023accelerating} predicts expert popularity before model deployment using the maximum a posteriori probability based on historical 
token-to-expert mapping and token ID information. 
Cong et al.~\cite{cong2024prediction} propose an LSTM-based algorithm trained on historical token-to-expert mapping for expert selection predication.
These expert selection prediction methods 
may achieve low prediction accuracy as the token-to-expert relationship 
in historical data is not exploited 
~\cite{nie2023flexmoe,wang2023prophet}, 
or are memory-intensive and require a long training time~\cite{cong2024prediction}. Lina~\cite{li2023accelerating} considers the token distribution 
but only uses token ID as the token feature.
Token ID is insufficient to fully identify a token in token-to-expert mappings, as tokens with the same token ID may be routed to different experts at a MoE layer, as illustrated in Fig.~\ref{fig_tid}. 

 \begin{figure}[!t]
 \centering
 \subfigure[256 tokens]{
   \includegraphics[width=0.23\textwidth]{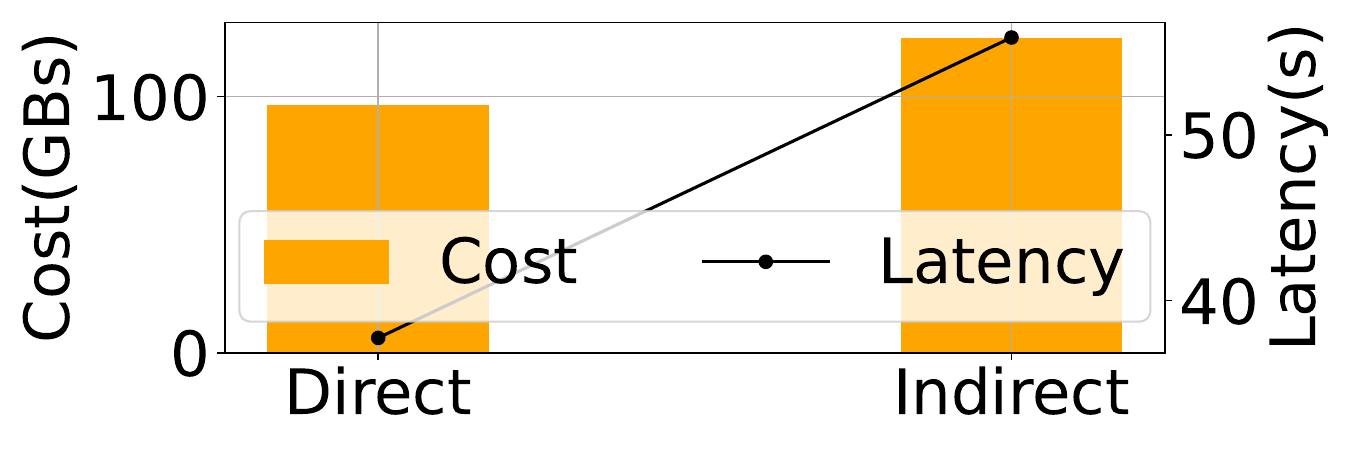}}
 \subfigure[2560 tokens]{
\includegraphics[width=0.23\textwidth]{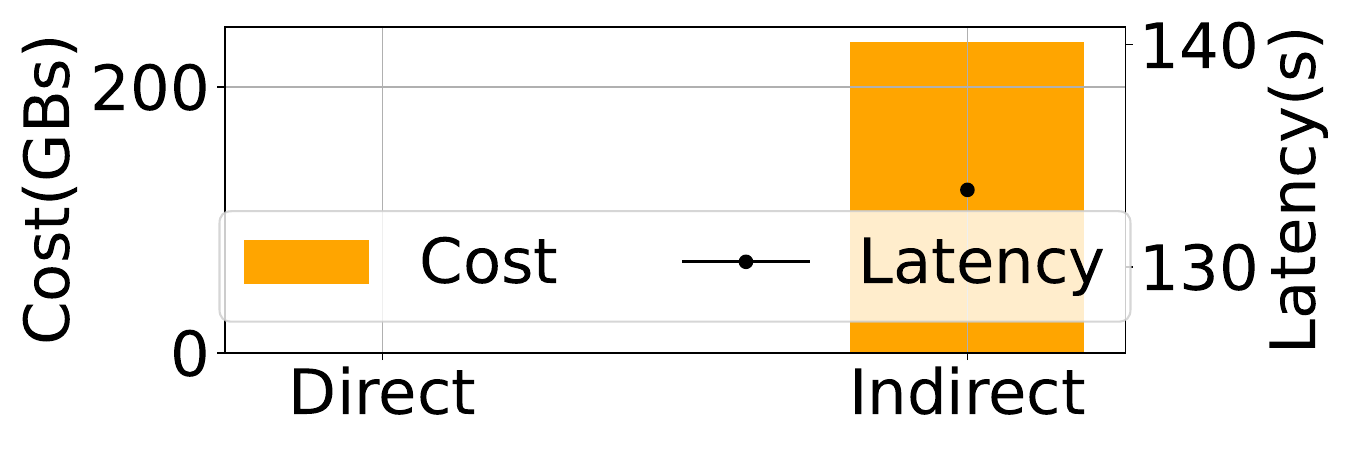}}
\vspace{-3mm}
 \caption{Billed cost of all MoE layers 
 and end-to-end inference time of a Bert-based MoE model on AWS Lambda (tokens from Enwiki8 dataset; 
 payload size 6MB 
 ).
 }
\label{motivation_alltoall}
\vspace{-5mm}
\end{figure}

\vspace{1mm}
\noindent\textbf{Challenge 2: MoE scatter-gather communication renders performance bottlenecks in serverless MoE inference.} 
In each MoE layer, scatter communication occurs between the gating network and expert networks, and gather communication between expert networks and the subsequent non-MoE layer,
which often renders a bottleneck  
because the non-MoE layer must wait for all experts to complete their computation and communication. 
The direct and indirect inter-function communication on a serverless platform results in different billed costs and inference time. 
In Fig.~\ref{motivation_alltoall}, MoE scatter-gather communication via indirect transfers incurs higher cost 
and longer inference time than direct transfers when serving a 256-token batch, due to the additional function running time required for storing data in external storage and retrieving data as input. 
Direct transfers cannot be adopted when serving a 2560-token batch as the payload size is exceeded; 
the serving cost under indirect transfers is very high. 
While pipelining communication with computation can typically alleviates this bottleneck in GPU/CPU clusters, 
pipelining in a serverless platform needs to take its direct or indirect data transfer modes into consideration, and carefully designed to improve efficiency in a serverless platform. 

We redesign token features and propose a novel posterior calculation method for expert selection prediction. 
We design and select the best scatter-gather communication methods for MoE layer inference.
We involve all modules in a BO framework for optimal MoE serving deployment.

\section{Design}\label{Secdesign}

\subsection{System overview}\label{systemoverview}
We consider distributed deployment of an MoE model in a serveless computing platform to serve inference requests. 
The MoE model consists of MoE layers (each includes a gating network and multiple experts) and non-MoE layers (e.g., feed-forward networks, multi-head attention networks \cite{fedus2022switch}). 
We adopt expert parallelism \cite{rajbhandari2022deepspeed, li2023accelerating} by assigning each expert to a serverless function. Model parallelism is adopted for non-MoE parts of the model with each non-MoE layer 
assigned to a serverless function. Model parameters are store in external storage.
We mainly focus on the MoE layers 
since the non-MoE layers are traditional DNNs extensively studied \cite{yu2021gillis} \cite{romero2021infaas}.

\begin{figure}[!t]
\centerline{\includegraphics[width=0.45\textwidth]{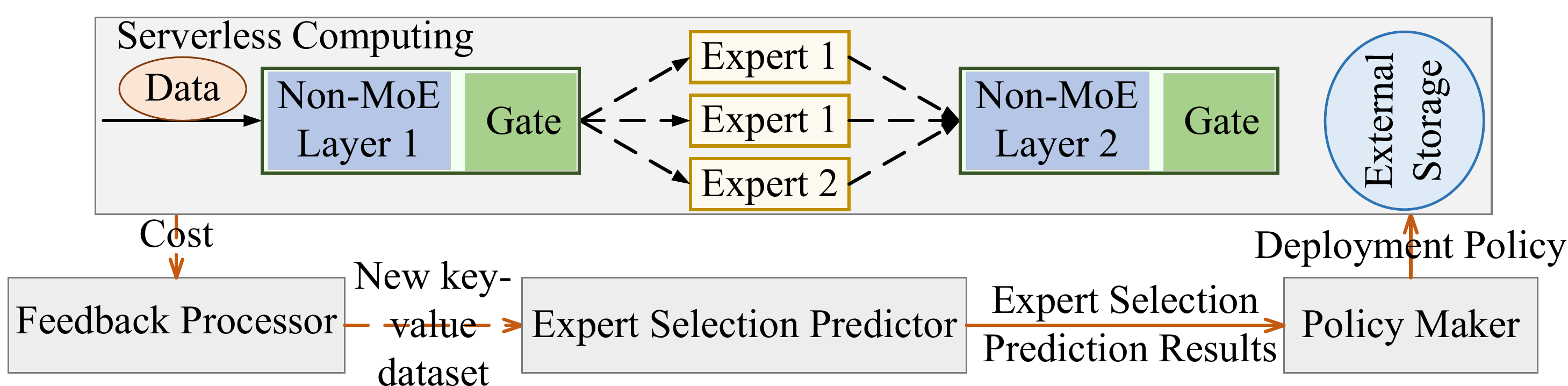}}
\vspace{-3mm}
\caption{System Overview. 
}
\label{sysoverview}
\vspace{-5mm}
\end{figure}

We propose a 
Bayesian Optimization (BO) framework to learn expert selections and optimal deployment of the MoE model for inference serving. The goal is to minimize overall billed cost of all MoE layers in serving. The BO framework consists of  
a {\em Feedback Processor} to adjust expert selection prediction, an {\em Expert Selection Predictor}, and a {\em Policy Maker}. 
An illustration is given in Fig.~\ref{sysoverview}. 
The expert selection predictor provides expert predictions of inference tokens from a real-world dataset on an inference task. This prediction is based on the posterior distribution calculated from these tokens and profiled data. The profiled data records the number of times each token-to-expert mapping occurs across at least 100 samples from the same real-world dataset, 
organized in a key-value dataset table. 
With the predictions, the policy maker decides how to deploy the MoE model, configures the memory size of each expert serverless function and adopts our scatter-gather communication design. 
The feedback processor adjusts the key-value pairs in the profiled dataset table for improving expert selection prediction, using feedback of the billed cost of all MoE layers in serving. In each BO iteration, expert predictor is adjusted, the policy maker decides optimal deployment of the MoE model, and the billed cost is collected for feedback processor's key-value table adjustment. 

When the BO algorithm converges, 
the MoE model is deployed according to the learned expert popularity and optimal deployment policy, and serves real-world inference requests.

\subsection{Expert selection prediction}\label{expertselection}


In an MoE model, each token in an inference request is routed to top-$k$ experts by the gating network at an MoE layer, based on token features. 
Token ID is the most commonly used token feature \cite{li2023accelerating}, but itself alone is insufficient to fully identify a token in token-to-expert mapping. We carefully design token features 
by investigating more token information during MoE inference. As Transformer models are typically the backbone of MoE models \cite{fedus2022switch}, we focus on Transformer-based MoE models.

For Transformer-based MoE models, token processing mainly occurs in the embedding, encoder and decoder layers. In the embedding layers, each token is embedded with its own information and its position (e.g., word embedding and position embedding). Thus, the token ID and position ID can be extracted as token features. In the encoder and decoder layers, tokens flow through feed-forward networks and one multi-head attention. In the feed-forward networks, no dependencies exist within the token sequence, so token ID and position ID are sufficient as the token features. 
Each multi-head attention layer contains multiple self-attentions. Each self-attention calculates the Query, Key, and Value of the tokens to capture dependencies between tokens in the token sequence.  The dependencies are quantified by the softmax attention scores, which indicate the relevance of each token to the others. For each token, we extract these dependencies as a token feature.
For simplicity, the dependencies for each token are derived as the token ID of the token with the highest sum of softmax attention scores across all self-attentions at a multi-head attention layer, referred to as the attention ID. 
The attention ID may vary before different MoE layers, aligning with the diverse expert popularities at different MoE layers. 
Therefore, the token features include the token ID, the position ID, and the attention ID. The token IDs and position IDs are from the input token sequences, known before inference starts. The attention IDs are from the self-attention parts of the non-MoE layer before each MoE layer, known during inference. 

Our expert prediction in the BO framework is learned on profiled data, which records the token-to-expert mappings on at least 100 samples from a real-word dataset of inference task. The profiled data are organized in a key-value dataset table where the keys represent token-to-expert mappings and the values denote their occurrence counts.
Especially, we design a new posterior calculation method and use the maximum posterior approach to predict expert selection for new tokens, where the posterior represents the probability of an expert given a token \cite{li2023accelerating}. 

Assume ${\bf f}$ is the token feature vector of a token, in which ${\bf f}_1$ is the token ID, ${\bf f}_2$ is the position ID and ${\bf f}_3$ is the attention ID. The posterior given the token can be expressed as $\mathcal{P}(\mathbb{N}_{e,i}|{\bf f})$ with $e\in\mathbb{E}$, where $\mathbb{N}_{e,i}$ is the $i$-th expert in the expert set $\mathbb{N}_{e}$ at MoE layer $e$ and $\mathbb{E}$ is the set of MoE layers in the MoE model. For a new token 
that has not undergone MoE inference, 
its feature ${\bf f}'_1$ is known but ${\bf f}'_3$ is unknown, 
The probability of ${\bf f}_2$ at any position is uniform, and the probability of ${\bf f}_3$ at any value can be approximated by the probability of ${\bf f}_1$ at that value, as the attention ID ${\bf f}_3$ is defined as the token ID with the highest attention scores. We can obtain all probabilities related to ${\bf f}'_1$ from the profiled data, 
the uniform probability $\mathcal{P}'({\bf f}_2)$ of ${\bf f}_2$ at any value and the probability $\mathcal{P}'({\bf f}_3)$ of ${\bf f}_3$ at any value from tokens in the same real-world dataset, that have not undergone MoE inference. 

To leverage all token features for identifying a token effectively, we use Bayes' theorem to design a new posterior calculation method. The Bayes' theorem $\mathcal{P}(\mathbb{N}_{e,i}|{\bf f})=\mathcal{P}({\bf f}|\mathbb{N}_{e,i})\mathcal{P}(\mathbb{N}_{e,i})/\mathcal{P}({\bf f})$ describes how the posterior of an expert selection given a token $\mathcal{P}(\mathbb{N}_{e,i}|{\bf f})$ is updated with the prior of the expert $\mathcal{P}(\mathbb{N}_{e,i})$, the prior of the token $\mathcal{P}({\bf f})$, and the likelihood of the token given the expert $\mathcal{P}({\bf f}|\mathbb{N}_{e,i})$. As $\mathcal{P}(\mathbb{N}_{e,i}|{\bf f}'_1)=\mathcal{P}({\bf f}'_1|\mathbb{N}_{e,i})\mathcal{P}(\mathbb{N}_{e,i})/\mathcal{P}({\bf f}'_1)$ and $\mathcal{P}({\bf f}'_1|\mathbb{N}_{e,i})=\mathcal{P}({\bf f}'_1,\mathbb{N}_{e,i})/\mathcal{P}(\mathbb{N}_{e,i})$, we can involve the uniform probability $\mathcal{P}'({\bf f}_2)$ of ${\bf f}_2$ and the probability $\mathcal{P}'({\bf f}_3)$ of ${\bf f}_3$ into the likelihood $\mathcal{P}({\bf f}'_1|\mathbb{N}_{e,i})$ of the expert given ${\bf f}'_1$, through the joint probability $\mathcal{P}({\bf f}'_1,\mathbb{N}_{e,i})$. For simplicity, we multiply $\mathcal{P}'({\bf f}_2)$ and $\mathcal{P}'({\bf f}_3)$ with $\mathcal{P}({\bf f}'_1,\mathbb{N}_{e,i})$ as involvement. Hence, the designed posterior calculation method is given by:
\begin{equation}\label{Byderive}
\small
\begin{aligned}
       \mathcal{P}(\mathbb{N}_{e,i}|{\bf f}'_1)
        = \int_{{\bf f}_{2}}&\int_{{\bf f}_{3}}\mathcal{P}^*(\mathbb{N}_{e,i}|{\bf f}'_1,{\bf f}_{2},{\bf f}_{3})\frac{\mathcal{P}^*({\bf f}'_1,{\bf f}_{2},{\bf f}_{3})\mathcal{P}'({\bf f}_{3})}{\mathcal{P}^{*}({\bf f}'_1,{\bf f}_{2})}d{\bf f}_{3}\\
        &\frac{\mathcal{P}^*({\bf f}'_1,{\bf f}_{2})\mathcal{P}'({{\bf f}_2})}{\mathcal{P}^{*}({\bf f}'_1)}d{\bf f}_{2}, \forall e \in \mathbb{E}, \forall i \in \mathbb{N}_e,
\end{aligned}
\end{equation}

where $\mathcal{P}^*(\cdot)$ represents the probability calculated from profiled data in the key-value dataset table.

Therefore, we can use the maximum a posterior method to predict the expert selection for a token with the known token feature ${\bf f}'_1$:

\vspace{-5mm}
\begin{equation}\label{ideaby}
    \small
    \hat{i}_{e}=argmax_{i\in\mathbb{N}_e} \mathcal{P}(\mathbb{N}_{e,i}|{\bf f}'_1), \forall e \in \mathbb{E},
\end{equation}
where $\hat{i}_{e}$ is the predicted expert at MoE layer $e$. Eq.~(\ref{ideaby}) can be readily extended to top-$k$ expert selection prediction.

\subsection{Scatter-gather communication design}\label{section2B}

We design pipelined scatter-gather communication at the MoE layer in a serverless platform for better resource utilization and cost reduction. 

For a batch of tokens to serve, we set a pipeline degree $\beta$ to split the batch for each expert into minibatches, 
where $\beta$ represents the maximal minibatch size. 
At each MoE layer, the gating network 
routes the splitted minibatches to each expert and the next non-MoE layer 
gathers processed minibatches from each expert.
If the minibatch size exceeds the payload limit, indirect transfer through external storage is used; otherwise, direct invocation of the serverless functions is adopted.

Data transfer from the gating network to experts can be pipelined with expert execution based on expert parallelism. Experts do not use pipelining for data transfer to the next non-MoE layer, as this may cause the next non-MoE layer to wait for the execution of experts (e.g., calculation of a minibatch) after receiving the previous processed minibatch. 
When external storage is involved, with pipelining, the time to download a minibatch from external storage and process this minibatch can overlap with the time to upload the previous processed minibatch to external storage.
Note that pipelining is only achievable with indirect transfer via external storage on a serverless platform. Serverless functions are stateless and direct data transfers from other functions require re-invocation of the function each time; model parameters that a serverless function uses are not retained during direct transfers and hence need to be reloaded from external storage for each re-invocation, resulting in significant time and memory waste.
We use one block time to represent the maximal overlap time of the indirect upload of a minibatch and the download and calculation of the next minibatch. 
The block time is determined by the pipeline degree $\beta$:
a larger $\beta$ results in fewer minibatches and longer block time. 

The benefits of pipelining can be reduced by the access delay to external storage. We design three scatter-gather communication methods tailored for MoE inference on a serverless platform. 
Let $a_e \in\mathbb{A}=\{1, 2, 3\}$ denote three possible scatter-gather communication methods at MoE layer $e$, and we allow all experts at an MoE layer to use the same method to simplify implementation.
Two indirect transfer options exist. 

\begin{figure}[!t]
 \centering
 \subfigure[With pipeline operation.]{
   \includegraphics[width=0.4\textwidth]{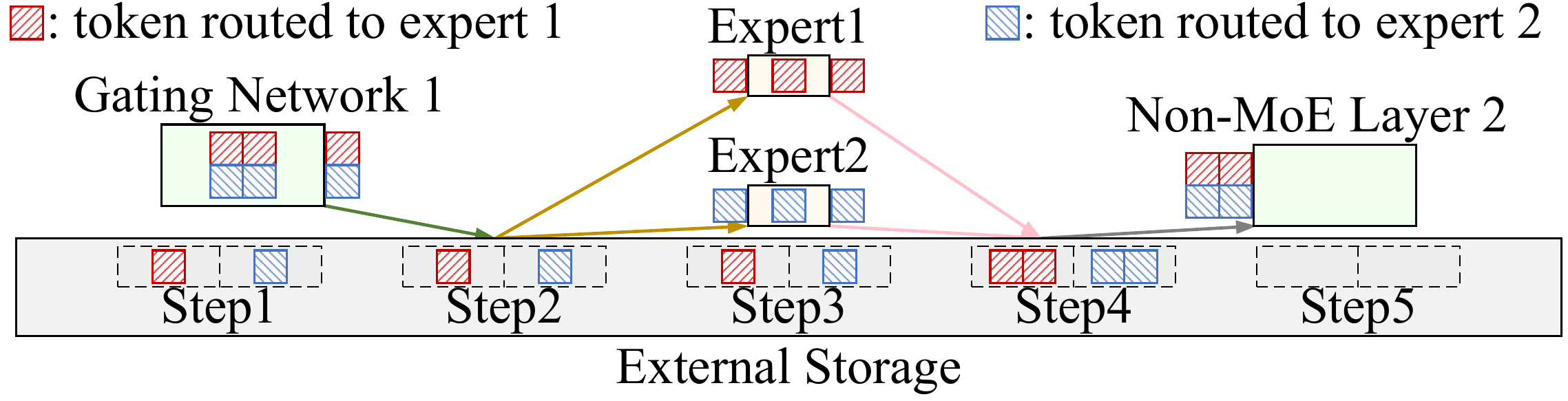}}
    \centering
 \subfigure[Without pipeline operation.]{
\includegraphics[width=0.4\textwidth]{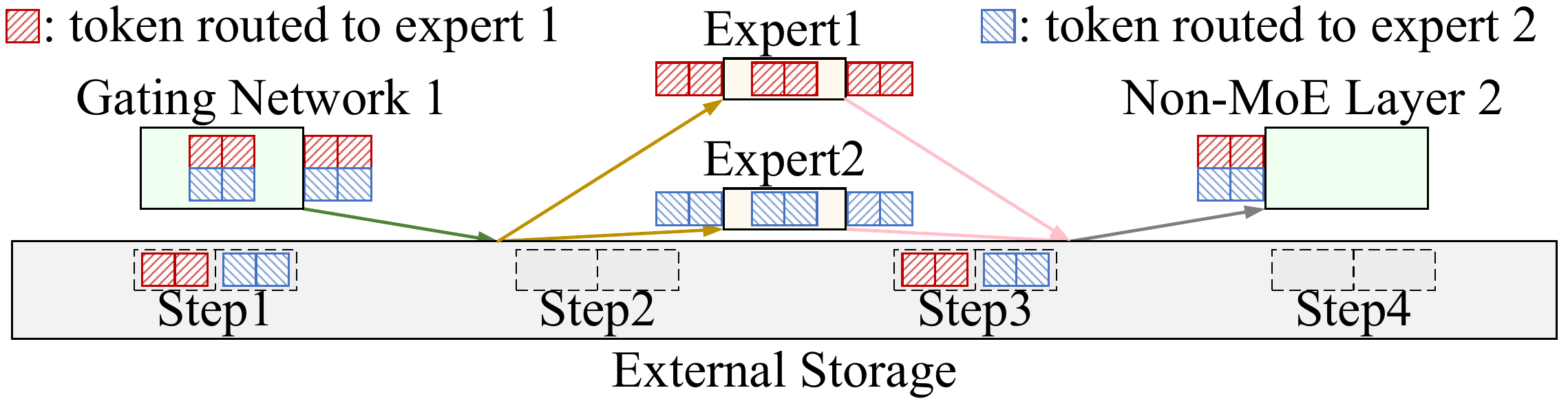}}
\vspace{-2mm}
 \caption{Scatter-gather communication 
 with indirect transfers through external storage: (a) with pipelining; (b) without pipelining. Pipeline degree $\beta$ is 2.
 }
\label{alltoallfig_indirect}
\vspace{-3mm}
\end{figure}

$\bullet$ In the first option ($a_e=1$), the gating network splits each expert's input into minibatches and sends them to external storage; at each expert, downloading a minibatch from external storage and calculating this minibatch can be overlapped with uploading the previous processed minibatch to external storage. 
As shown in Fig.~\ref{alltoallfig_indirect}(a), after splitting input data into minibatches of one token each, two minibatches for two experts are stored in external storage (step 1); these two minibatches are downloaded from external storage and calculated by two expert serverless functions, while the next two minibatches are stored in external storage (step 2); the processed minibatches are stored in external storage, while the next two minibatches are downloaded from external storage and calculated by two expert serverless functions (step 3); all processed minibatches are stored in external storage (step 4) and then the next non-MoE layer downloads all minibatches from external storage (step 5). 

$\bullet$ In the second option ($a_e=2$), the gating network transfers each expert's input via external storage to each expert function, and all processed results are downloaded by the next non-MoE layer, without pipelining. As illustrated in Fig.~\ref{alltoallfig_indirect}(b), all input data are stored in external storage (step 1); each expert serverless function downloads its input from external storage (step 2), and stores the processed results in external storage (step 3); then the next non-MoE layer downloads all processed results from external storage (step 4). 

\begin{figure}[!t]
 \centering
 \subfigure[With pipeline operation.]{
   \includegraphics[width=0.23\textwidth]{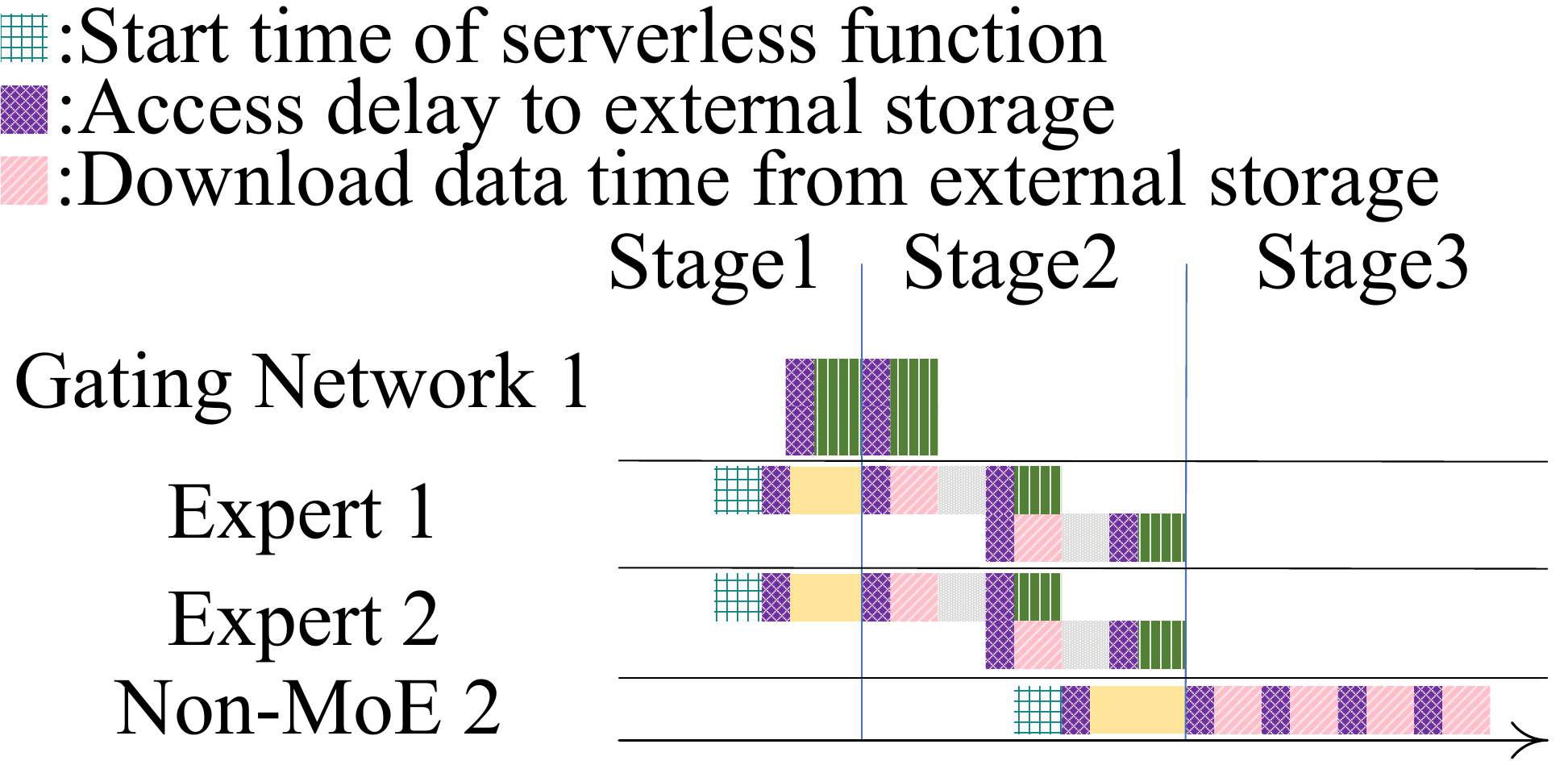}}
 \subfigure[Without pipeline operation.]{
   \includegraphics[width=0.23\textwidth]{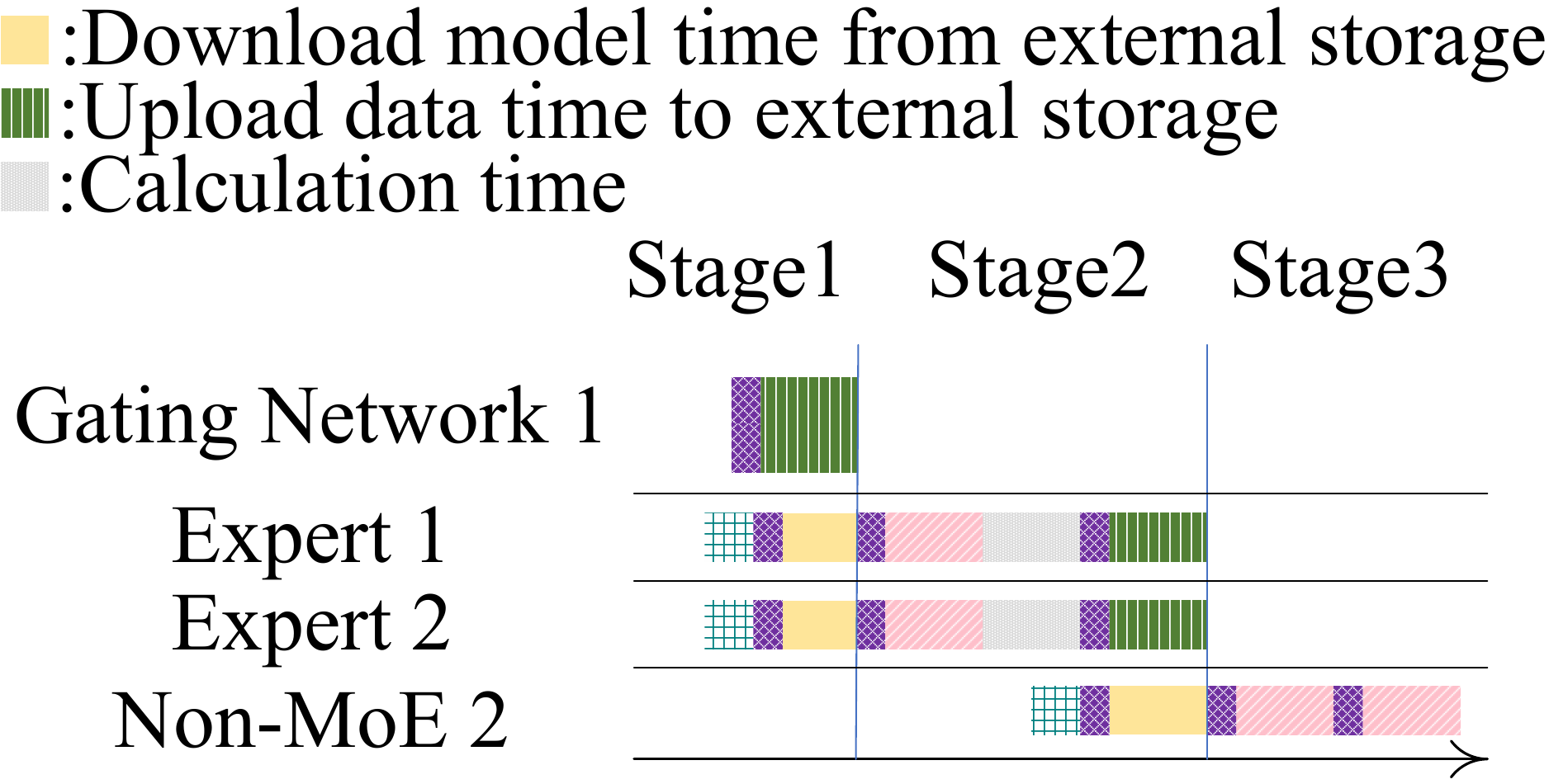}}
   \vspace{-2mm}
 \caption{Scatter-gather communication time 
 with indirect transfers through external storage: (a) with pipelining; (b) without pipelining.
 }
\label{alltoalltimeblockfig_indirect}
\end{figure}

Fig.~\ref{alltoalltimeblockfig_indirect} illustrates the execution time of MoE layers under the two communication designs. 
Stage 1 in both cases represents the time for each expert to start the warm-up function without resource initialization (i.e., short warm start time) and download model parameters, while the gating network concurrently uploads each expert's input to external storage. Stage 2 is the time for each expert to download input from external storage and calculate it, overlapping with the time to start the warm-up function and download model parameters in the next non-MoE layer. In stage 3, the next non-MoE layer downloads the processed results. 

\begin{figure}[!t]
 \centering
\includegraphics[width=0.4\textwidth]{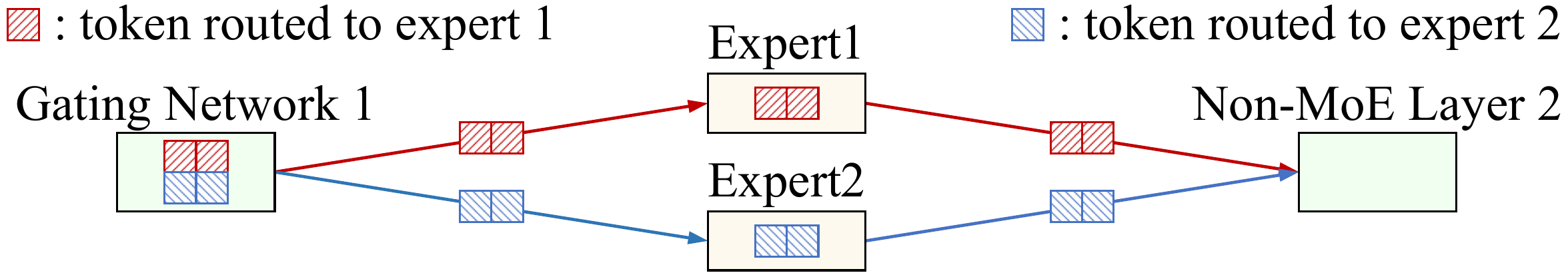}
\vspace{-3mm}
 \caption{Scatter-gather communication 
 with direct function invocation.
 }
\label{alltoallfig_direct}
\vspace{-3mm}
\end{figure}

\begin{figure}[!t]
 \centering
\includegraphics[width=0.43\textwidth]{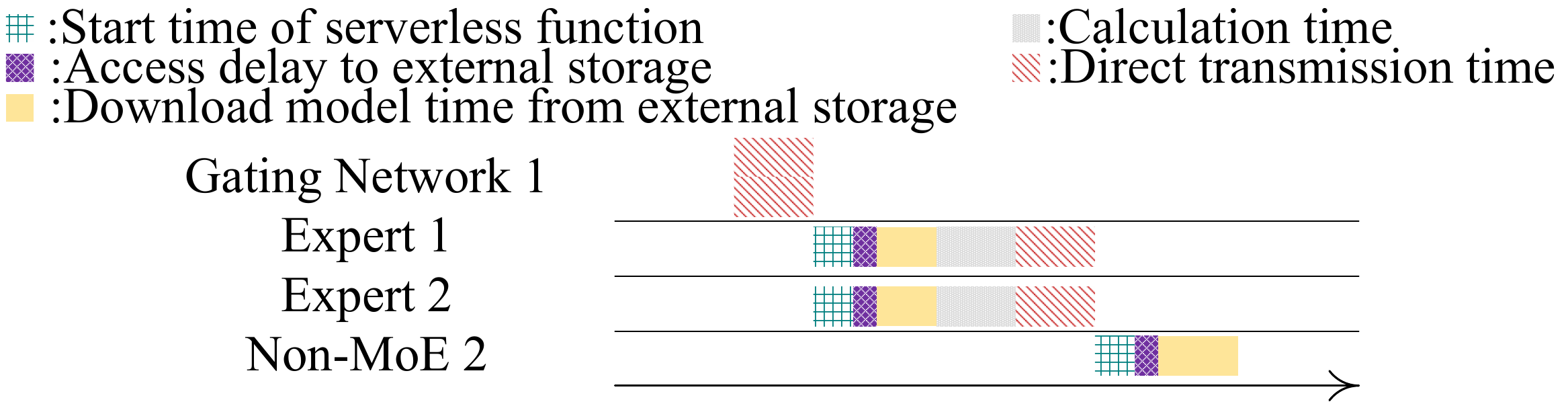}
\vspace{-3mm}
 \caption{Scatter-gather communication time 
 with direct function invocation. 
 }
\label{alltoalltimeblockfig_direct}
\vspace{-5mm}
\end{figure}

$\bullet$ When using direct transfers ($a_e=3$), the gating network directly transfers each expert's input to the expert serverless function, and the processed results of each expert are directly transferred to the next non-MoE layer, as shown in Fig.~\ref{alltoallfig_direct}. Fig.~\ref{alltoalltimeblockfig_direct} illustrates its execution time. 

The total MoE layer time varies with different communication designs. 
We will carefully make the choices in our distributed MoE deployment problem. 

\subsection{MoE model deployment
}\label{Deployformulation}

The policy maker decides optimal MoE model deployment by making the following decisions:

$\bullet$ Memory size configuration for each serverless function. Let $\mathbb{M}$ be the set of memory size options for each serverless function (e.g., from 128MB to 3008MB on AWS Lambda)
$x_{e,i,j}\in\{0,1\}$ denotes if the $j$-th option in set $\mathbb{M}$ is selected for expert $i$ in MoE layer $e$ 
(1) or not (0). 
The processing time of one token at expert $i$ in MoE layer $e$, 
is given by:

\vspace{-5mm}
\begin{equation}
\small
t^{cal}_{e,i}=s_{e,i}\sum^{|\mathbb{M}|}_{j=1}x_{e,i,j}U_j,
        \forall e\in \mathbb{E},\forall i \in \mathbb{N}_e,
\end{equation}
\vspace{-2mm}

\noindent where $s_{e,i}\in\{0,1\}$ denotes if the expert is selected (1) or not (0) for the token (given based on expert selection prediction), 
and $U_j$ is the time to process one token in an expert
using the $j$-th memory size option with the serverless function. 

$\bullet$ Expert replication. As the maximal memory size of each serverless function is limited, it is possible that it still takes a long time for a popular expert to process tokens routed to it. Given that an end-to-end inference time target should be met in inference serving, we further consider replicating serverless functions of experts and allow expert replicas to run in parallel for toke processing.
Let $y_{e,i,g}\in\{0,1\}$ denote if expert $i$ in MoE layer $e$ has $g$ 
serverless function replicas (1) or not (0), where $g=1, ..., G$ with $G$ as maximal possible replica number
The number of tokens routed to one replica is 
$r_{ e,i}=\sum^{G}_{g=1}y_{e,i,g}d_{e,i}/g$, where $d_{e,i}$ denotes the number of tokens routed to all replicas of the expert.
Let $D^{in}$ be the size of one token and $D^p$ be the maximal payload size in the serverless platform. 
When $r_{e,i} D^{in}>D^p$, direct transfer 
is not feasible (i.e, ${a_e}=1$ or ${a_e}=2$) at MoE layer $e$. 

$\bullet$ Scatter-gather communication method and parameter ($\beta$). We seek to minimize the total billed cost of all MoE layers. 
The billed cost of the gating network can be ignored here, as it affects the cost of all MoE layers little: the memory size of serverless functions for gating networks does not depend on expert popularity, and the impact of scatter-gather communication methods on the gating network's execution time is also reflected in the experts.
Hence, we consider the total billed cost of all MoE layers to be the billed cost of all experts in MoE layers. 
The billed cost of MoE layer $e$ is then given by $c_e=(a_e-2)(a_e-3)c_{1,e}+(a_e-1)(a_e-3)c_{2,e}+(a_e-1)(a_e-2)c_{3,e}$, where the billed cost of MoE layer $e$ under the communication method $a_e$ is: 
\vspace{-1mm}
\begin{equation}
    \small
    c_{a_e,e}=\sum_{i\in \mathbb{N}_e}s_{e,i}t_{a_e,e,i}\sum^M_{j=1}x_{e,i,j}\mathbb{M}_j,\forall e\in\mathbb{E},\forall a_e\in\mathbb{A},
\end{equation}
Here $\mathbb{M}_j$ represents the $j$-th memory size in set $\mathbb{M}$, and 
$t_{a_e,e,i}$ is the total execution time of all replicas of the expert:
\vspace{-1mm}
\begin{equation}
    \small
t_{a_e,e,i}=\sum^{G}_{g=1}y_{e,i,g} g t^{rep}_{a_e,e,i},\forall e\in \mathbb{E},\forall i \in \mathbb{N}_e,\forall a_e\in\mathbb{A},\label{firstconst}
\end{equation}
where $t^{rep}_{a_e,e,i}$ denotes the execution time of one replica of the expert. 
$t^{rep}_{a_e,e,i}$ is related to the scatter-gather communication method chosen. 
We give formulars of three cases in obtaining $t^{rep}_{a_e,e,i}$.

\vspace{1mm}
\noindent\textit{(1) Pipelined indirect transfer 
($a_e=1$).} 
We have 
\begin{equation}
    \small
t^{rep}_{1,e,i}=T^{h,E}_{e,i}+t^{nblk}_{e,i}+\beta t^{blk}_{e,i},\forall e\in \mathbb{E},\forall i \in \mathbb{N}_e,\label{eq_t1rep}
\end{equation}
where 
the head time $T^{h,E}_{e,i}=\frac{P_{e,i}}{B^s}+T^{dl}+T^{str}$ consists of warm start time $T^{str}$, the access delay to external storage $T^{dl}$ and the model download time $\frac{P_{e,i}}{B^s}+T^{dl}$ with bandwidth $B^s$ between external storage and serverless function and the parameter size $P_{e,i}$ of the expert. 
$t^{nblk}_{e,i}=T^{dl}+\lceil\frac{r_{e,i}}{\beta}\rceil(\frac{D^{o}}{B^s})$ includes the time to upload the last minibatch to external storage 
and $D^{o}$ is the size of the processed result of one token by an expert. 
$t^{blk}_{1,e,i}=T^{dl}+\beta\max\{\frac{D^{in}}{B^s}+t^{cal}_{e,i},\frac{D^{o}}{B^s}\}$ denotes one worst-case block time. 

The latency, 
from the earlist time point when expert serverless functions start or the gating network starts to transfer each expert's input, to the latest time when the next non-MoE layer finishes downloading the processed results of all experts from external storage or finishes downloading model parameters with direct transfer, is referred to as MoE-E2E latency 
$t^{lat}_{1,e}$: 
\begin{equation}
    \small
      t^{lat}_{1,e}=\max\{t^{S12}_{1,e},T^{load}_e\}+t^{S3}_{1,e},\forall e\in \mathbb{E},\label{eq_tlat1}
\end{equation}
where 
$T^{load}_e$ includes the time to start the serverless function of non-MoE layer and download the model parameters. 

These formulars are derived based on Fig.~\ref{alltoalltimeblockfig_indirect}(a) and details omitted due to space limit.

\vspace{1mm}
\noindent\textit{(2) Non-pipelined indirect transfer ($a_e=2$).} We have 
\begin{equation}
    \small
t^{rep}_{2,e,i}=T^{h,E}_{e,i}+2T^{dl}+t^{data}_{2,e,i},
\forall e\in \mathbb{E},\forall i \in \mathbb{N}_e,\label{Eq_rep2}
\end{equation}
where $t^{data}_{2,e,i}=r_{e,i}(\frac{D^{in}+D^{o}}{B^s}+t^{cal}_{e,i})$.  

The MoE-E2E latency is: 
\begin{equation}
    \small
       t^{lat}_{2,e}=\max\{t^{S12}_{2,e},T^{load}_e\}+t^{S3}_{2,e},\forall e\in \mathbb{E}. \label{eq_lat2}
\end{equation}

These formulars are derived based on Fig.~\ref{alltoalltimeblockfig_indirect}(b) and details omitted due to space limit

\vspace{1mm}
\noindent \textit{(3) Direct transfer ($a_e=3$).} We have: 
\vspace{-1mm}
\begin{equation}\label{eq_rep3}
    \small
        t^{rep}_{3,e,i}=T^{h,E}_{e,i}+r_{e,i}(\frac{D^{o}}{B^f}+t^{cal}_{e,i}),\forall e\in \mathbb{E},\forall i \in \mathbb{N}_e,
        \vspace{-1mm}
\end{equation}
where $B^f$ is the bandwidth between serverless functions.

The MoE-E2E latency $t^{lat}_{3,e}$ is: 
\vspace{-1mm}
\begin{equation}
    \small
        t^{lat}_{3,e}=\frac{r_{e,i}D^{in}}{B^f}+\max_{ i \in \mathbb{N}_e}\{t^{rep}_{3,e,i}\}+T^{load}_e,\forall e\in \mathbb{E}.\label{eq_lat3}
        \vspace{-1mm}
\end{equation}

The MoE-E2E latency at MoE layer $e$ is given by $t^{lat}_{e}=(a_e-2)(a_e-3)t^{lat}_{1,e}+(a_e-1)(a_e-3)t^{lat}_{2,e}+(a_e-1)(a_e-2)t^{lat}_{3,e}$.




\noindent\textbf{Optimal MoE deployment problem:} We formulate optimal deployment of MoE model inference in a serverless platform to minimize the billed cost of all MoE layers (i.e., $\sum_{e\in\mathbb{E}}c_{e}$), by jointly deciding the communication method (i.e., $a_e$), selecting memory size configurations (i.e. $x_{e,i})$, deciding expert replication (i.e. $y_{e,i}$), and setting parameter $\beta$ for pipelined scatter-gather communication.

\begin{subequations}\label{formulation_v1}
\small
  \begin{align}
  &\min\sum_{e\in\mathbb{E}}c_{e} \label{obj}\\
     &\text{subject to }\quad\ref{firstconst}-\ref{eq_lat3} \label{formulation_v1_b},\\
&P_{e,i}+M^{itrm}_{e,i}+r_{e,i}(D^{in}+D^{o})\nonumber\\
&\qquad \leq\sum^{|\mathbb{M}|}_{j=1}x_{e,i,j}\mathbb{M}_j, \forall e \in \mathbb{E},\forall i\in \mathbb{N}_e\allowdisplaybreaks\label{memory_limit}\\
&T^{head}+T^{tail}+\sum_{e\in \mathbb{E}}(t^{lat}_{e}+T^{NE}_e)\leq T^{limit},\label{time_limit}\\
&1\leq\beta\leq \max_{i\in \mathbb{N}_e, e\in \mathbb{E}}r_{e,i},\allowdisplaybreaks\label{beta_limit}\\
&(a_e-3)(r_{e,i}D^{in}-D^p)\leq 0,\forall e\in \mathbb{E},\forall i\in \mathbb{N}_e,\forall a_e\in\mathbb{A}, \allowdisplaybreaks\label{direct_limit}\\
&\sum^{|\mathbb{M}|}_{j=1}x_{e,i,j}=1,\forall e\in \mathbb{E},\forall i\in \mathbb{N}_e,\allowdisplaybreaks\\
&\sum^{G}_{g=1}y_{e,i,g}=1,\forall e\in \mathbb{E},\forall i\in \mathbb{N}_e,\allowdisplaybreaks\\
&x_{e,i,j}\in\{0,1\}, \forall e \in \mathbb{E},\forall i\in \mathbb{N}_e,\forall j \in \mathbb{M},\allowdisplaybreaks\\
&y_{e,i,g}\in\{0,1\},\forall e \in \mathbb{E},\forall i\in \mathbb{N}_e,  \forall g \in \mathbb{G}_{e,i},\allowdisplaybreaks\\
&\beta\in \mathbb{Z},
  \end{align}
\end{subequations}
Here $M^{itrm}_{e,i}$ is the memory size of intermediate results during an expert's inference.  
$T^{limit}$ is the time limit of end-to-end MoE model inference (i.e., serving SLO). 
$T^{head}$ is the execution time of the serverless function of the first non-MoE layer. 
$T^{tail}$ is the execution time of the serverless function of the last non-MoE layer, excluding the time to receive data from the last MoE layer or download them from external storage. 
$T^{NE}_e$ is the processing time 
of non-MoE layer $e$ with the subsequent gating network. 
(\ref{memory_limit}) specifies the memory limit of each serverless function, (\ref{time_limit}) gives the end-to-end inference time target of the MoE model,
(\ref{beta_limit}) limits the maximal number of tokens in each block in calculating the worst-case total time, and (\ref{direct_limit}) prohibits direct tranfers when the payload size is below transferred data size between the gating network and each expert, as well as between each expert and the next non-MoE layer.

\section{The BO Framework}\label{secAlg}
We now present our BO algorithm for the BO framework to learn expert selection predictions and optimize MoE model deployment, together with an efficient algorithm to solve the optimal MoE model deployment by the policy maker. 

\subsection{Optimal MoE Deployment Algorithm}
Given expert selection results, the optimal MoE deployment problem in (\ref{formulation_v1}) 
is a MIQCP problem with a non-linear objective in (\ref{obj}), quadratical constraints in  (\ref{memory_limit}) and (\ref{direct_limit}), binary variables $x$ and $y$, and integer variable $\beta$ and $a$. MIQCP problems are in general NP-hard \cite{elloumi2019global} because of their non-linearity and 
discretized variables. 
We design an efficient algorithm to solve it approximately. 

We first divide problem~(\ref{formulation_v1}) into three cases by fixing the scatter-gather communication method: $a_e=1$,$\forall e\in\mathbb{E}$; $a_e=2$,$\forall e\in\mathbb{E}$; and $a_e=3$,$\forall e\in\mathbb{E}$. We then linearize max functions in (\ref{obj}), (\ref{time_limit}) and (\ref{beta_limit}) 
by adding auxilliary variables  
($\max_{h\in \mathbb{H}}\{h\}$ can be
linearized as $\phi\geq h$, $\forall h \in \mathbb{H}$
). 
Then we solve each resulting MIQCP by a solver \cite{Gurobi}, respectively, and obtain costs $c_{1,e}$, $\forall e\in\mathbb{E}$, $c_{2,e}$, $\forall e\in\mathbb{E}$, and $c_{3,e}$, $\forall e\in\mathbb{E}$, from the three solutions. 
Based on these solutions, we design an Optimal Deployment Selection (ODS) algorithm to decide $a_e$ for each MoE layer, as given in Alg~\ref{alg_optimal_deploy}. 

For each MoE layer $e$, we select the communication method $\hat{a}_e$ with the lowest cost, set all experts to use the same method, 
and then calculate the MoE-E2E latency $\hat{t}^{lat}_{\hat{a}_e,e}$ (lines 4-7). If the new MoE-E2E latency satisfies the end-to-end inference time constraint, the optimal deployment policy for the MoE model is obtained; otherwise, we identify the MoE layer $\tilde{e}$ with the highest latency, set the cost of the corresponding scatter-gather communication $\tilde{a}_{\tilde{e}}$ to infinity (lines 10-12), and then iteratively decide $\hat{a}_e$ for each MoE layer (lines 3-17). At most $2|\mathbb{E}|$ iterations are needed, as three communication methods provide up to $3|\mathbb{E}|$ solutions of $\hat{a}_e$, $\forall e\in\mathbb{E}$, and selecting $|\mathbb{E}|$ solutions excludes up to $2|\mathbb{E}|$ other solutions.
If all costs $c_{a_e,e}$ become infinity, 
it implies that mixing different communication methods across different MoE layers do not work. In this case, we return the optimal deployment policy with the lowest cost with all MoE layers using the same scatter-gather communication method (lines 18-20).

\subsection{BO algorithm}


\begin{algorithm}[!t]
\small
   \caption{Optimal Deployment Selection Algorithm of an MoE model (ODS) 
   }
    \label{alg_optimal_deploy}
    \begin{algorithmic}[1]
    
        \REQUIRE Optimal solutions $x_{a,e,i}$, $y_{a,e,i}$, $\beta$ and optimal objective $c_{a,e}$,$\forall e\in\mathbb{E}$,$\forall i\in\mathbb{N}_e$ of solving three MIQCP problems with fixed $a$,$\forall a \in \mathbb{A}$
        \ENSURE $\hat{a}_e$, $\beta$, $x_{\hat{a}_e,e,i}$,$y_{\hat{a}_e,e,i}$,$\forall e\in\mathbb{E}$,$\forall i\in\mathbb{N}_e$
        \STATE itr=0.
        \WHILE{
        itr$\leq2|\mathbb{E}|$}
        \FOR{$e\in\mathbb{E}$ 
        }
        \FOR{$i\in\mathbb{N}_e$}
        \STATE $\hat{a}_e=argmin_{a_e\in\mathbb{A}}[c_{1,e},c_{2,e},c_{3,e}]$; 
        \STATE Calculate the MoE-E2E latency $\hat{t}^{lat}_{\hat{a}_e,e}$.
        \ENDFOR
        \ENDFOR
        \IF{$T^{head}+T^{tail}+\sum_{e\in\mathbb{E}}(\hat{t}^{lat}_{\hat{a}_e,e}+T^{NE}_e)>T^{limit}$}
        \STATE $\tilde{a}_{\tilde{e}},\tilde{e}=argmin_{e\in\mathbb{E}}\hat{t}^{lat}_{\hat{a}_e,e}$;
        \STATE Set $c_{\tilde{a}_{\tilde{e}},\tilde{e}}=\infty$.
        \ELSE \RETURN $\hat{a}_e$, $x_{\hat{a}_e,e,i}$, $y_{\hat{a}_e,e,i}$, $\beta$,$\forall e\in\mathbb{E}$,$\forall i\in\mathbb{N}_e$.
        \ENDIF
        \STATE itr+=1.
        \ENDWHILE
        \IF{itr$>2|\mathbb{E}|$}
        \STATE $\hat{a}=argmin_{a\in\mathbb{A}}\sum_{e\in\mathbb{E}}c_{a,e}$
        \RETURN $\hat{a}$, $x_{\hat{a},e,i}$, $y_{\hat{a},e,i}$, $\beta$,$\forall e\in\mathbb{E}$,$\forall i\in\mathbb{N}_e$.
        \ENDIF
    \end{algorithmic}
\end{algorithm}

\begin{algorithm}[!th]
\small
   \caption{Bayesian Optimization Algorithm with Multiple-dimension ${\bf \epsilon}$-Greedy Search 
   }
    \label{alg1}
    \begin{algorithmic}[1]
        \REQUIRE MoE model, dataset table $\Omega_0$, 
        normal range of key-value pairs $\mathbb{P}$,
        and constants $Q$, $\mu$, $\alpha$, $\rho$, $\rho_1$, $\rho_2$, $\rho_3$, $\lambda$, $\zeta$.
        \ENSURE optimal key-value pairs $\{\hat{{\bf z}}_{q},\hat{v}_{q}\}_{q\in\{1,\dots,Q\}}$
        \STATE Initialize Q key-value pairs $\{{\bf z}_{0,q}, v_{0,q}\}_{\forall q \in\{1,\dots,Q\}}$ for key-value table, ${\bf \epsilon}_0\in\mathbb{R}^{Q}$ for ${\bf \epsilon}$-GS, limited range of key-value pairs $\mathbb{L}$, BO historical set 
        $\mathbb{B}_0=\{\}$, and the BO trial index $\tau=1$.
        \REPEAT
        \STATE ${\bf \epsilon}_{\tau}=\frac{{\bf \epsilon}_0}{1+\rho \tau}$.
        \STATE $\Omega_\tau({\bf z}_{\tau-1,q})=v_{\tau-1,q}$,$\forall q \in\{1,\dots,Q\}$.
        \STATE $\hat{i}_{e}=argmax_{i\in\mathbb{N}_e} \mathcal{P}(\mathbb{N}_{e,i}|{\bf f}'_1)$,$\forall e \in \mathbb{E}$ with $\Omega_{\tau}$.
        \STATE $c_{\tau,a,e}$ from three MIQCP solvers,$\forall e\in\mathbb{E},\forall a\in\mathbb{A}$.
        \STATE $(\hat{a}_e, x_{\hat{a}_e,e,i},y_{\hat{a}_e,e,i},\beta)_{\tau}$=ODS($c_{\tau,a,e}$),$\forall a\in\mathbb{A}$, $\forall e\in\mathbb{E}$, $\forall i\in\mathbb{N}_e$.
        \FOR{$j=1,\dots,J$}
        \FOR{$e\in\mathbb{E}$}
        \FOR{$i\in\mathbb{N}_e$}
        \IF{$|r_{e,i}- R^{real}_{e,i}|>\alpha$ 
        }
        \STATE Append ${\bf f}'_{j,1}$ to $\mathbb{L}_{\tau}$.
        \IF{$M^{real}\geq \sum^{|\mathbb{M}|}_{j=1}x_{\tau,\hat{a}_e,e,i,j}{\mathbb{M}}_j$ 
        }
        \STATE $\rho'=\rho_1$, $n^{new}_{e,i}=\lceil \frac{M^{real}}{\sum^{|\mathbb{M}|}_{j=1}x_{\tau,\hat{a}_e,e,i,j}\mathbb{M}_j}\rceil$. 
        \ELSIF{$\hat{a}_{\tau,e}=3$ and $R^{real}_{e,i}>D^p$}
        \STATE $\rho'=\rho_2$, $n^{new}_{e,i}=\lceil \frac{R^{real}_{e,i}}{D^p}\rceil$.
        \ELSE
        \STATE $\rho'=\rho_3$, $n^{new}_{e,i}=1$.
        \ENDIF
        \STATE ${\bf \epsilon}_{\tau,1:\mu Q}=(1+\rho'\tau){\bf \epsilon}_{\tau,1:\mu Q}$.
        \STATE Replicate 
        expert $i$ $n^{new}_{e,i}$ times.
        \ENDIF
        \ENDFOR
        \ENDFOR
        \STATE
        Deploy 
        the MoE model using $(\hat{a}_e, x_{\hat{a}_e,e,i},y_{\hat{a}_e,e,i},\beta)_{\tau}$;
        \STATE $c_{\tau,j}=MoE_{\tau}(input_{j})$. 
        \ENDFOR
        \STATE
        $c_{\tau}=\frac{1}{J}\sum^J_{j=1}c_{\tau,j}$.
        \STATE $\mathbb{B}_{\tau}\leftarrow\mathbb{B}_{\tau-1}$$\cup(\{{\bf z}_{\tau-1,q}, v_{\tau-1,q}\}_{q\in\{1,\dots,Q\}},c_{\tau})$.
        \STATE ${\bf z}_{\tau,1:\mu Q},v_{\tau,1:\mu Q}\leftarrow {\bf \epsilon}_{\tau,1:\mu Q}-$GS over $\mathbb{B}_{\tau}$, $\mathbb{L}_{\tau}$.
        \STATE ${\bf z}_{\tau,\mu Q+1:Q},v_{\tau,\mu Q+1:Q}\leftarrow {\bf \epsilon}_{\tau,1:\mu Q+1:Q}-$GS over $\mathbb{B}_{\tau}$, $\mathbb{P}$.
        \STATE $\tau=\tau+1$, initialize $\mathbb{L}$.
        \UNTIL{The change of $\min_{\tau\in{1,\dots,\tau}}c_{\tau}$ within $\lambda$ consecutive iterations is below $\zeta$, and record the last iteration index as $\eta$}
        \RETURN$\{\hat{{\bf z}}_{q},\hat{v}_{q}\}_{q\in\{1,\dots,Q\}}=argmin_{\tau\in\{1,\dots,\eta\}}{c_{\tau}}$.
    \end{algorithmic}
\end{algorithm}


BO is a statistical approach for global optimization of a black-box function, including an objective, variables, a surrogate function to simulate the objective, and an acquisition method to update the variables. BO makes no assumptions about the underlying function format and aims to minimize the number of trials to find a near-optimal solution, reducing tuning costs. 
In our BO algorithm, the black box function corresponds to the billed cost of all MoE layers for MoE inference in a serverless platform. 
Each trial corresponds to one BO iteration which adjusts the key-value table, recomputes expert selection predictions and distributed MoE model deployment. 

The objective of our BO is to minimize the billed cost of all MoE layers. 
The variables are $Q$ key-value pairs in the key-value table $\Omega(\cdot)$ for adjusting expert selection prediction. The surrogate function uses a Gaussian process to simulate the cost of all MoE layers deployed by the policy maker based on expert selection prediction.
For the acquisition method, we design a decaying multi-dimensional ${\bf \epsilon}$-greedy search (GS) to set the variables for the next BO iteration. Traditional 
single-dimension $\epsilon$-GS, as an acquisition function, decays with each iteration and balances exploration of BO variables and exploitation of high-performing BO variable values  
by selecting the best variable 
with probability $1-\epsilon$ and exploring new variable values with probability $\epsilon$ \cite{de2020epsilon}. 
However, as multiple key-value pairs in key-value table need to be adjusted together in a single BO trial, these pairs can be viewed as multi-dimensional variables in our BO, which make 
a single-dimension $\epsilon$ insufficient to balance exploration and exploitation in all dimensions. We extend $\epsilon$ to a multi-dimensional vector ${\bf \epsilon}\in\mathbb{R}^Q$. 

In BO learning, our objective, billed cost of all MoE layers, is obtained on several batches of inference data from an open-source real-world dataset \cite{enwik8,lambda,wmt19} (different datasets can be used for different MoE inference tasks). 
When inaccurate expert selection prediction occurs compared to the ground truth in the profiled data, 
we set a limited range of key-value pairs to update as $\mathbb{L}$, where $\mathbb{L}$ includes all positive integers for values and all possible token-to-expert mappings with token IDs limited to those present in these batches of data for keys. 
 We set a ratio $\mu$, and slow down the decay of the split ${\bf \epsilon}_{1:\mu Q}$ of vector ${\bf \epsilon}$ from the first dimension to the $\mu Q$-th dimension by multiplying ${\bf \epsilon}_{1:\mu Q}$ with a factor greater than 1. 
The 1st to $\mu Q$-th key-value pairs in our key-value table 
are then updated using ${\bf \epsilon}_{1:\mu Q}$ by adjusting the values of keys in $\mathbb{L}$, 
allowing for more exploration on current low-performing key-value pairs. 
Meanwhile, the $\mu Q + 1$-th through the last key-value pairs in the variables are updated by either adjusting the values of keys in $\mathbb{P}$ or by creating new key-value pairs in $\mathbb{P}$ using ${\bf \epsilon}_{\mu Q+1: Q}$. Here, $\mathbb{P}$ is a normal range of key-value pairs to adjust, includeing all positive integers for values, and all possible token-to-expert mappings for keys, where token features can include all possible token IDs assigned by the tokenizer, all possible position IDs, and all possible attention IDs, 
and experts are represented by all possible expert indices.
This enriches the key-value table by creating new key-value pairs with keys as token-to-expert mappings not present in the profiled data.

The BO algorithm with multi-dimensional ${\bf \epsilon}$-GS 
is given in Alg.~\ref{alg1}. In the $\tau$-th BO iteration, 
${\bf \epsilon}$ decays by being divided by $1+\rho \tau$ with a constant $\rho>0$ (line 3). The dataset table $\Omega_{\tau}$ is updated with key-value pairs $\{{\bf z}_{\tau-1,q}, v_{\tau-1,q}\}_{\forall q \in\{1,\dots,Q\}}$ (line 4). 
where ${\bf z}=\{{\bf f}, e, i\}$. 
Then expert selection is predicted (line 5). The policy maker produces the optimal deployment policy using 
the ODS algorithm (lines 6-7).
For the $j$-th batch $input_j$ in BO learning
(lines 8-27), at expert $i$ in MoE layer $e$, if predicted counts $r^{e,i}$ and real counts $R^{real}_{e,i}$ of tokens assigned to one replica of this expert
exceeds a constant $\alpha>0$, token IDs ${\bf f}'_{j,0}$ of the $j$-th batch are recorded for limiting the range of key-value pairs to adjust 
(line 12) and three cases are discussed (lines 10-21): (i) if the minimal memory $M^{real}$ required by real expert popularity
exceeds memory configuration of serverless functions, 
$\rho_1<\rho$ is used to decrease the decay rate ${\bf \epsilon}_{\tau, 1:\mu Q}$, and $n^{new}_{e,i}$ is calculated to replicate expert $i$ $n^{new}_{e,i}$ times to satisfy the minimal memory $M^{real}$ 
(lines 13-14, 20-21); (ii) if the size of transferred tokens exceeds the payload size under direct transfer, $\rho_2<\rho_1$ is used to decrease the decay rate ${\bf \epsilon}_{\tau,1:\mu Q}$, and $n^{new}_{e,i}$ is calculated to replicate expert $i$ $n^{new}_{e,i}$ times to ensure that the data size transferred to each replica does not exceed the payload size $D^p$
(lines 15-16, 20-21); (iii) if all constraints in~(\ref{formulation_v1}) are satisfied, 
$\rho_3<\rho_2$ is used to decrease the decay rate ${\bf \epsilon}_{\tau, 1:\mu Q}$, and we do not replicate expert $i$ to avoid increasing cost 
(lines 17-18, 20-21). Then 
the billed cost of all MoE layers $c_{\tau,j}$ is computed on the $j$-th batch data $input_j$ on the derived MoE model deployment $MoE_{\tau}$ (lines 25, 26, 28).
Next the historical set $\mathbb{B}$ in BO to record variables and objectives 
is updated (line 29), and the key-value pairs to adjust is updated by ${\bf \epsilon}$-GS over the historical set 
and the range of key-value pairs $\mathbb{P}$ and $\mathbb{L} $
(lines 30-31). 
Then BO iterations repeat until the change of the minimal billed cost of all MoE layers within $\lambda$ consecutive iterations is below the threshold $\zeta$ (line 33). 


\subsection{Theoretical Analysis} 


\noindent\textbf{Theorem 1.} \textit{Alg.~\ref{alg_optimal_deploy} produces feasible MoE deployment in $O(|\mathbb{E}|)$ time, which achieves a billed cost of all MoE layers upper bounded by a constant ratio of the cost of optimal solutions of (\ref{formulation_v1}).}

\noindent\textbf{Theorem 2.} 
\textit{Alg.~\ref{alg1} converges when the BO iteration index satisfies $\tau>\frac{1+\rho}{\rho-\rho_1}(1-\frac{\delta}{\max_{q\in\{1,\dots,Q\}}\epsilon_{0,q}})$ with an abitrary small positive constant $\delta<\max_{q\in\{1,\dots,Q\}\epsilon_{0,q}}$.
}

Theorem 1 indicates that the time complexity of Alg.~\ref{alg_optimal_deploy} scales linearly with the number of MoE layers. 
Theorem 2 shows that the Alg.~\ref{alg1} converges in a constant bound. 
The sketch of proods are as follows.

\noindent\textit{Sketch of Proof 1:} In theorem 1, the time complexity is analyzed in this section. We prove that the billed cost by the optimal solutions \textbf{OPT} of (\ref{formulation_v1}) is lower-bounded by the ideal \textbf{OPT\_LB}. as $\sum_{e\in\mathbb{E}}\min_{a_e\in\mathbb{A}}c_{a_e,e}$.
On the other hand, the billed cost achieved by Alg.~\ref{alg_optimal_deploy} (i.e., \textbf{ALG}) is upper bounded by \textbf{ALG\_UP} as $\sum_{e\in\mathbb{E}}\max_{a_e\in\mathbb{A}}c_{a_e,e}$. 
From Sec~\ref{Deployformulation}, $c_{a_e,e}>\sum_{i\in\mathbb{N}_e}\{T^{h,E}_{e,i}\}+\sum_{i\in\mathbb{N}_e}\{d_{e,i}\}(U_{|\mathbb{M}|}+\min\{1/B_s,1/B_f\})$, and $c_{a_e,e}<\mathbb{M}_{|\mathbb{M}|}G(\sum_{i\in\mathbb{N}_e}\{T^{h,E}_{e,i}\}+\sum_{i\in\mathbb{N}_e}\{d_{e,i}\}(U_{1}+\max\{1/B_s,1/B_f\}+T^{dl})$. Here the known head time $T^{h,E}_{e,i}$ as the first term, the number of tokens multiply the known unit calculation and transfer time as the second term, and $\mathbb{M}_{|\mathbb{M}|}$ and $G$ denotes the known maximal memory size and number of replicas. Then \textbf{ALG}$/$\textbf{OPT}$\leq$\textbf{ALG\_UP}$/$\textbf{OPT\_LB}$\leq$$\mathbb{M}_{|\mathbb{M}|}G(U_{1}+\max\{1/B_s,$$1/B_f\}+T^{dl})$$/(U_{|\mathbb{M}|}+\min\{1/B_s,1/B_f\})$. Proved.

\noindent\textit{Sketch of Proof 2:} In theorem 2, when ${\bf \epsilon}$ decays below an arbitrary small positive constant $\delta<\max_{q\in\{1,\dots,Q\}\epsilon_{0,q}}$, we consider the variables achieving the best objective in BO historical set are always selected by GS. As ${\bf \epsilon}$ decays at the slowest rate $(1+\rho_1\tau)/(1+\rho\tau)$, we set $\max_{q\in\{1,\dots,Q\}\epsilon_{0,q}}(1+\rho_1\tau)/(1+\rho\tau)<\delta$. Proved. 


\section{Evaluation}\label{Secevaluation}
\subsection{Experimental Setup}
\noindent\textbf{Testbed.} We run our experiments on AWS Lambda \cite{Amazon}. 
To build MoE layer images, 
we use a Dockerfile to define the environment with Python 3.8 and include packages such as \emph{torch} and \emph{transformers}. 
We implement the BO algorithm with package {\em optuna} \cite{optuna} and MIQCP solvers with package {\em gurobi} \cite{Gurobi}.
We use two S3 buckets of size 512MB each for external storage. 
We adopt 14 discrete memory size configurations for each serverless function: [128, 768, 960, 1152, 1344, 1536, 1728, 1920, 2112, 2304, 2496, 2688, 2880, 3072] MB. 
We set maximal possible expert replica number as 8.

\vspace{1mm}
\noindent\textbf{MoE Models.} Three common transformer-based dense language models are converted to MoE models with all MLP layers after attention layers converted to MoE layers and a gating network of a linear layer: 

$\bullet$ Bert \cite{devlin2018bert}: a 12-layer encoder model with 110 million parameters, converted to 12 MoE layers, with each MoE layer having 4, 8, or 16 experts;

$\bullet$ GPT2 \cite{radford2019language}: a 12-layer decoder model with 1.5 billion parameters, converted to 12 MoE layers, with each MoE layer having 4 experts;

$\bullet$ Bert2Bert \cite{chen2021bert2bert}: a 12-layer encoder-decoder model with 247 million parameters, converted to 24 MoE layers, with each MoE layer having 4 experts.

We run the fill-mask task \cite{fillmask} on Enwik8 \cite{enwik8} and CCnews \cite{ccnews} datasets and the translation task \cite{translation} on the Wmt19 \cite{wmt19} dataset for the BERT model. We conduct the text generation task on the Enwik8 and Lambda \cite{lambda} datasets for the GPT-2 model and on the Enwik8 dataset for Bert2Bert model.

 \begin{figure}[!t]
 \centering
\includegraphics[width=0.5\textwidth]{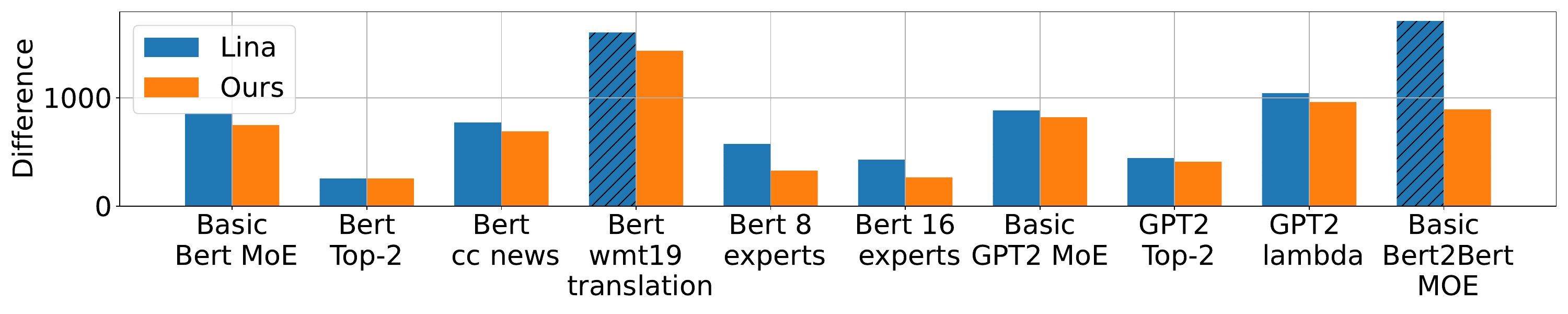}
\vspace{-5mm}
 \caption{Average difference per expert between real and predicted expert selection numbers under different MoE models, datasets and tasks.
 }\label{fig_expertselection}
\vspace{-5mm}
\end{figure}

 \begin{figure}[!t]
 \centering
 \subfigure[Bert cost.]{
   \includegraphics[width=0.23\textwidth]{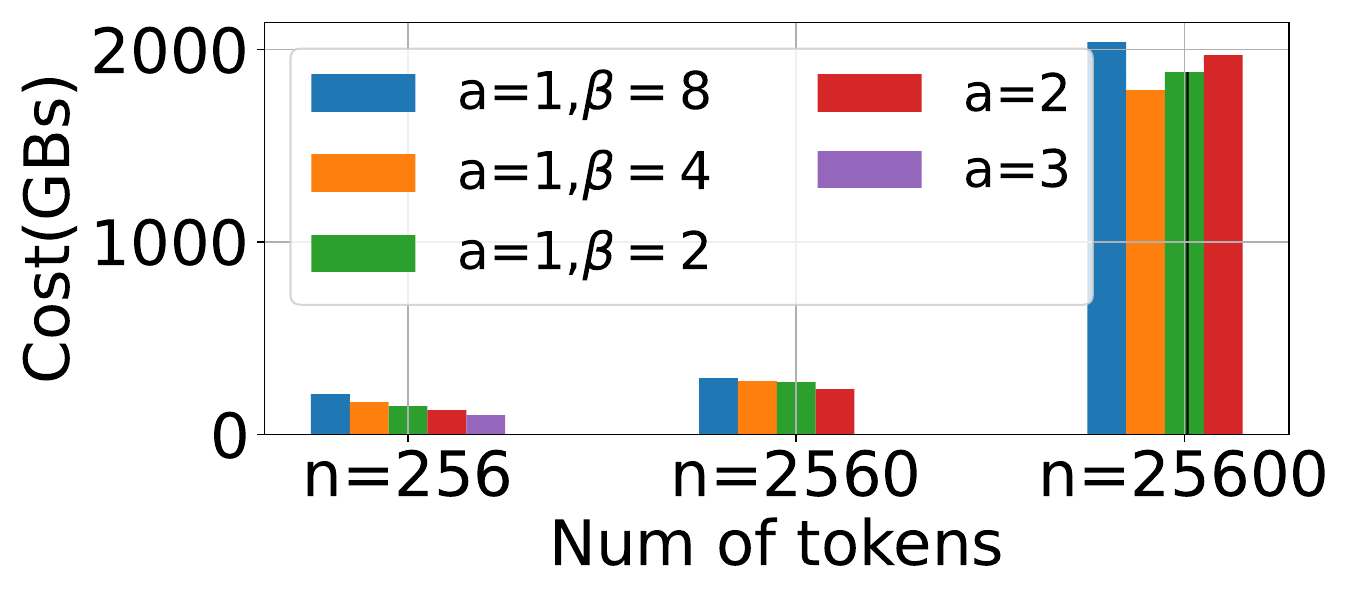}}
 \subfigure[Bert throughput.]{
   \includegraphics[width=0.23\textwidth]{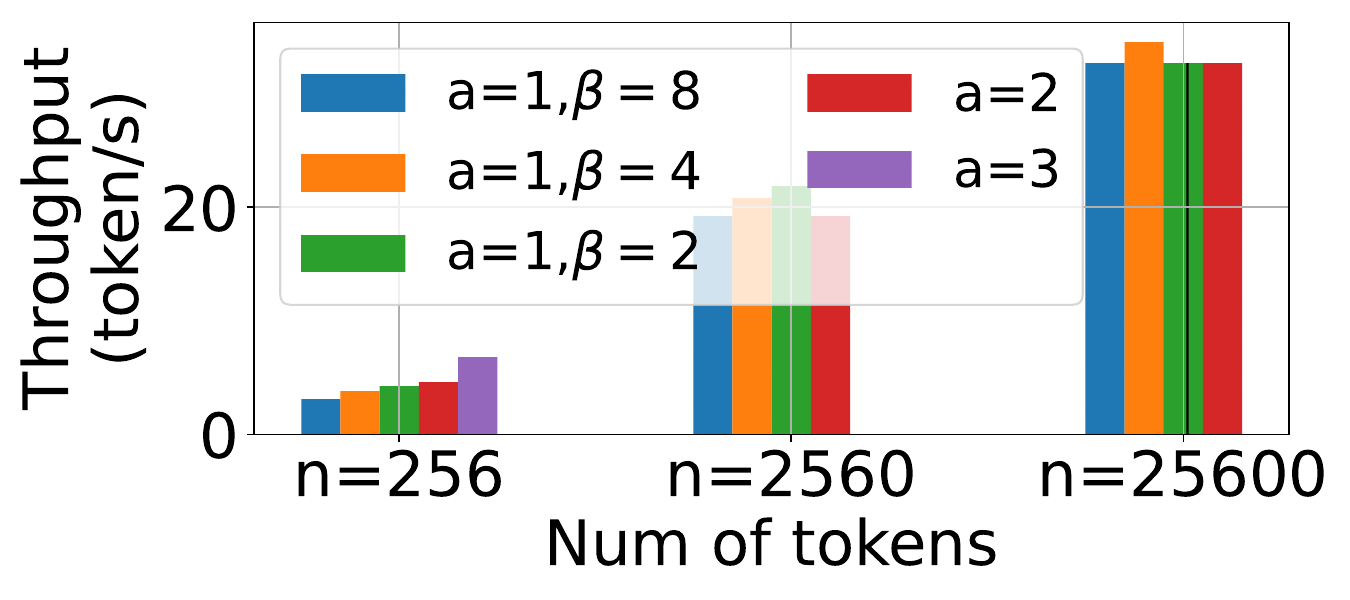}}
   \subfigure[GPT2 cost.]{
   \includegraphics[width=0.23\textwidth]{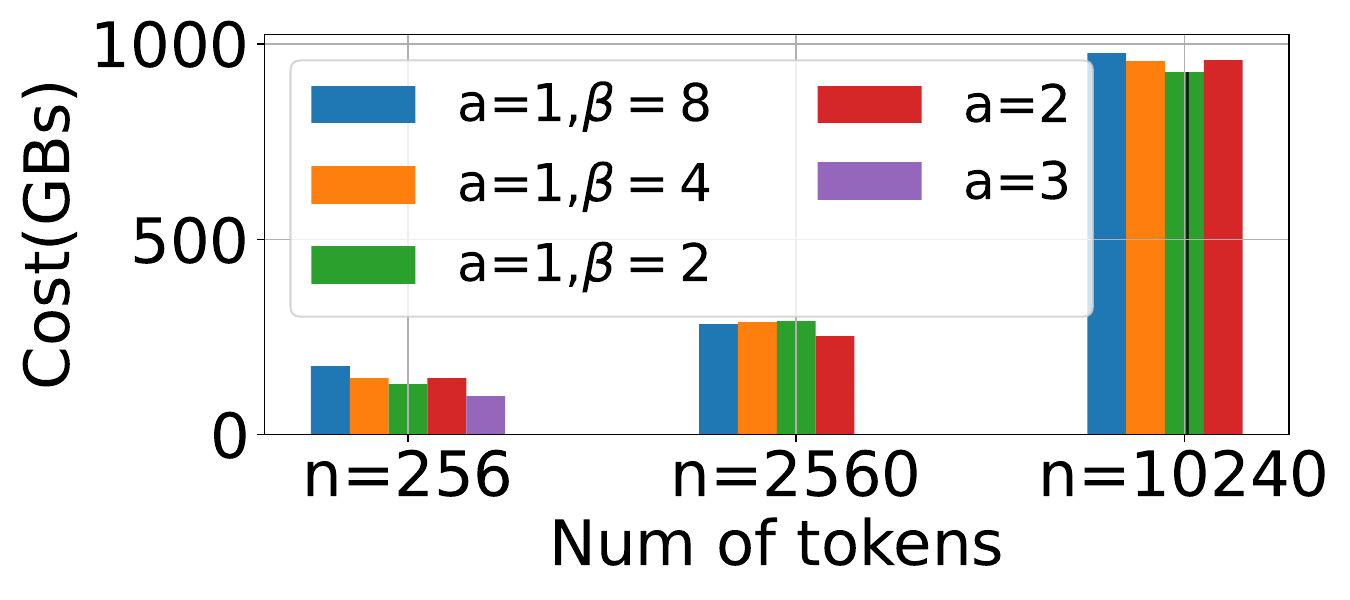}}
 \subfigure[GPT2 throughput.]{
   \includegraphics[width=0.23\textwidth]{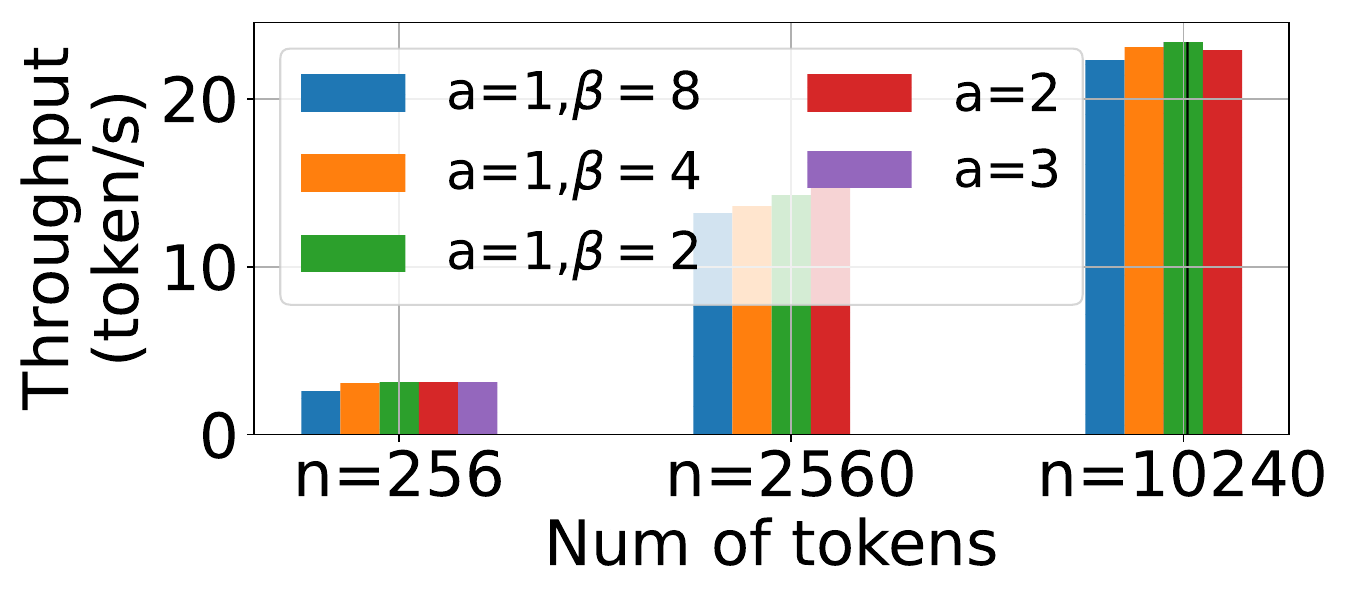}}
   \vspace{-3mm}
 \caption{Billed cost and inference throughput of MoE layers with different scatter-gather communication methods 
 on AWS Lambda.}\label{fig_comm}
 \vspace{-3mm}
\end{figure}

 \begin{figure}[!t]
 \centering
 \subfigure[Bert.]{
   \includegraphics[width=0.23\textwidth]{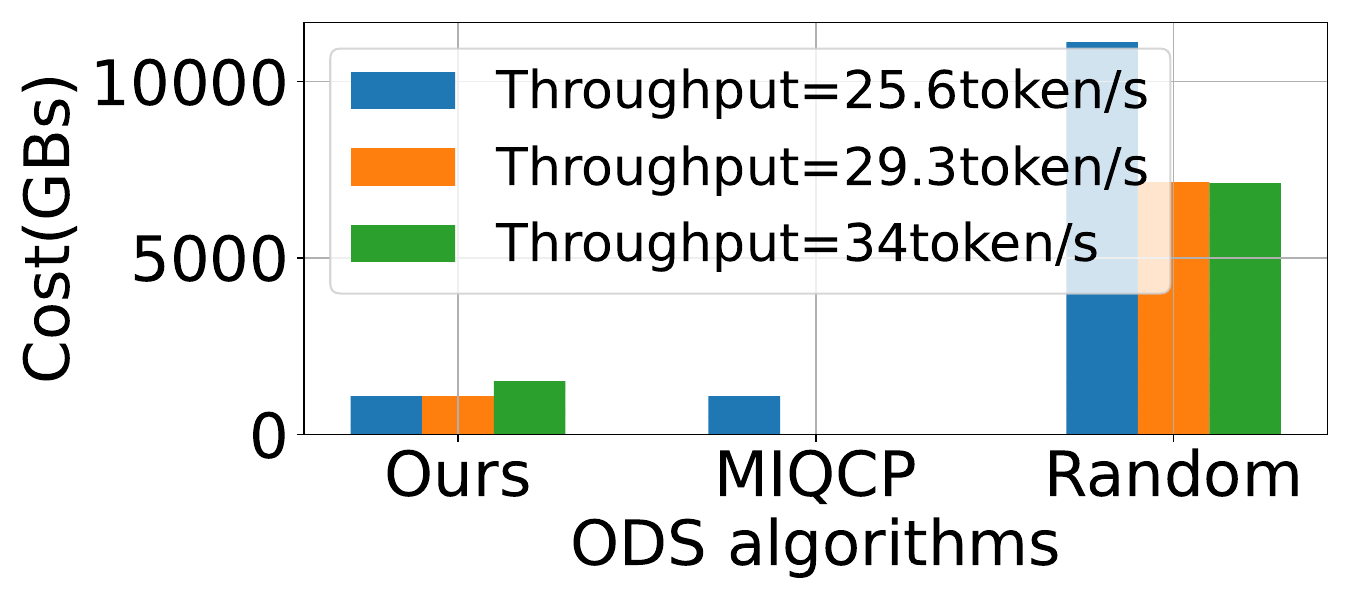}}
   \subfigure[GPT2.]{
   \includegraphics[width=0.23\textwidth]{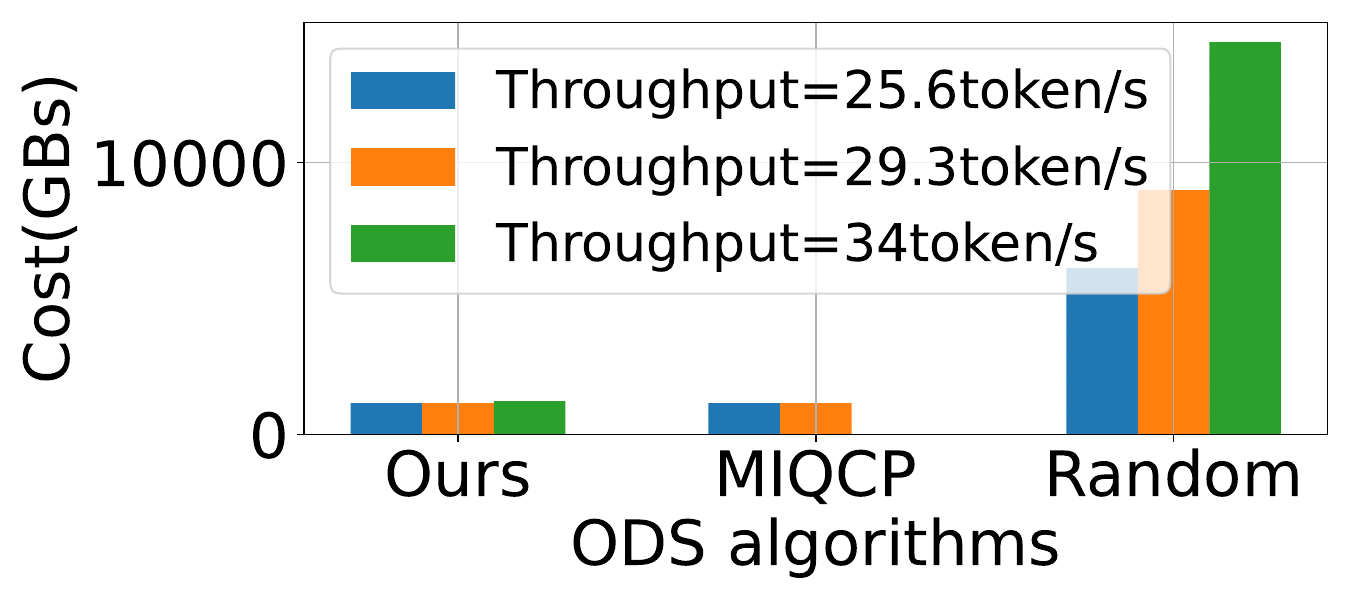}}
   \vspace{-3mm}
 \caption{Billed cost of all MoE layers with different MoE deployment algorithms on AWS Lambda. 
 }\label{fig_ODS}
\end{figure}

 \begin{figure}[!t]
 \centering
 \subfigure[Bert cost.]{
   \includegraphics[width=0.23\textwidth]{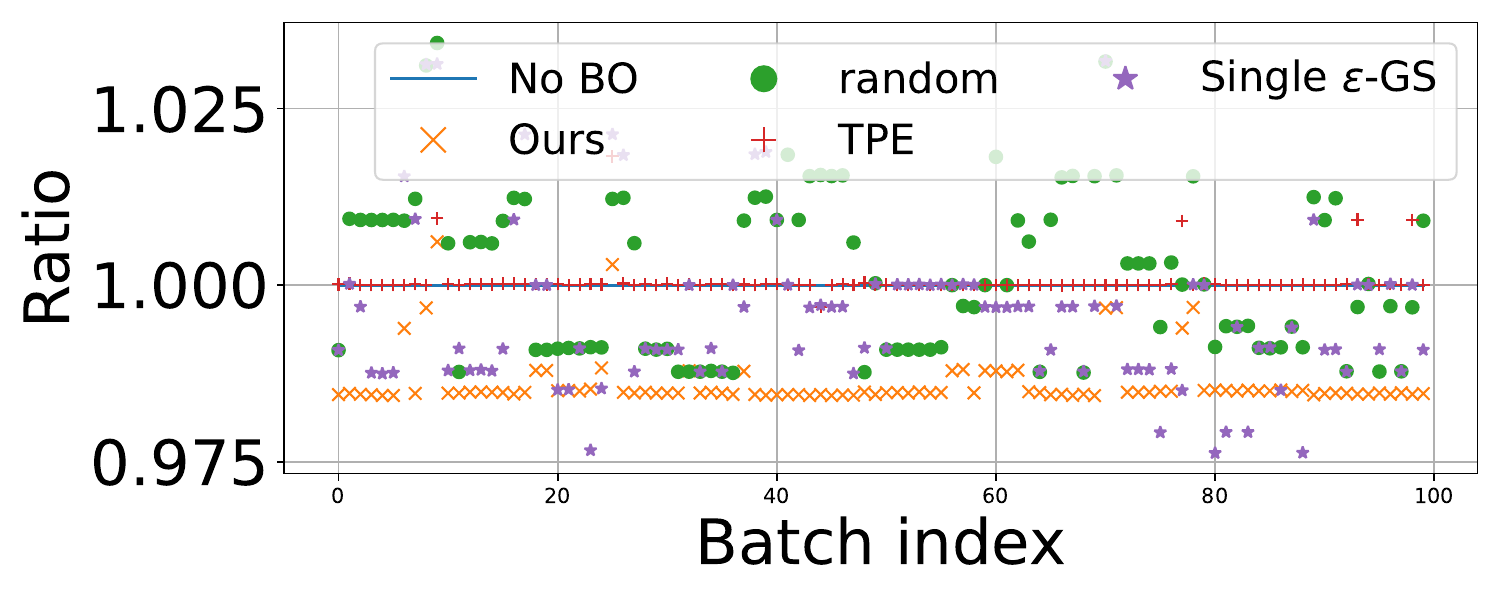}}
 \subfigure[Bert difference.]{
   \includegraphics[width=0.23\textwidth]{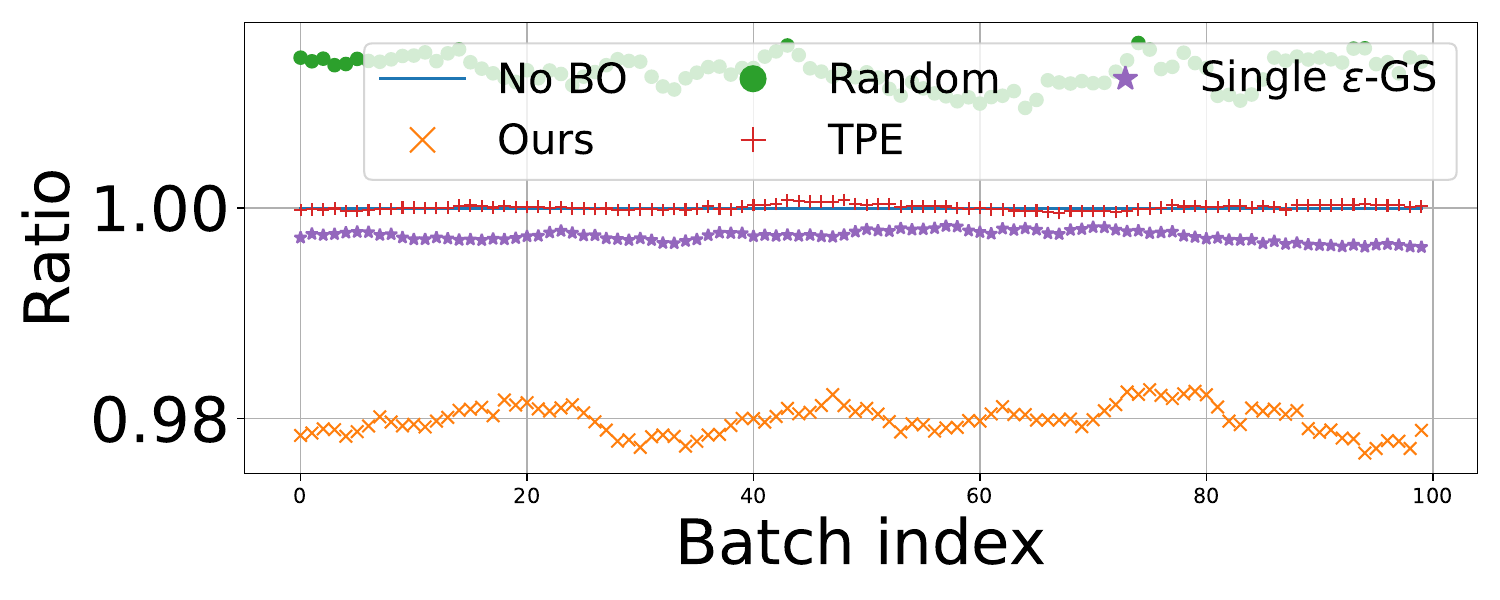}}
   \subfigure[GPT2 cost.]{
   \includegraphics[width=0.23\textwidth]{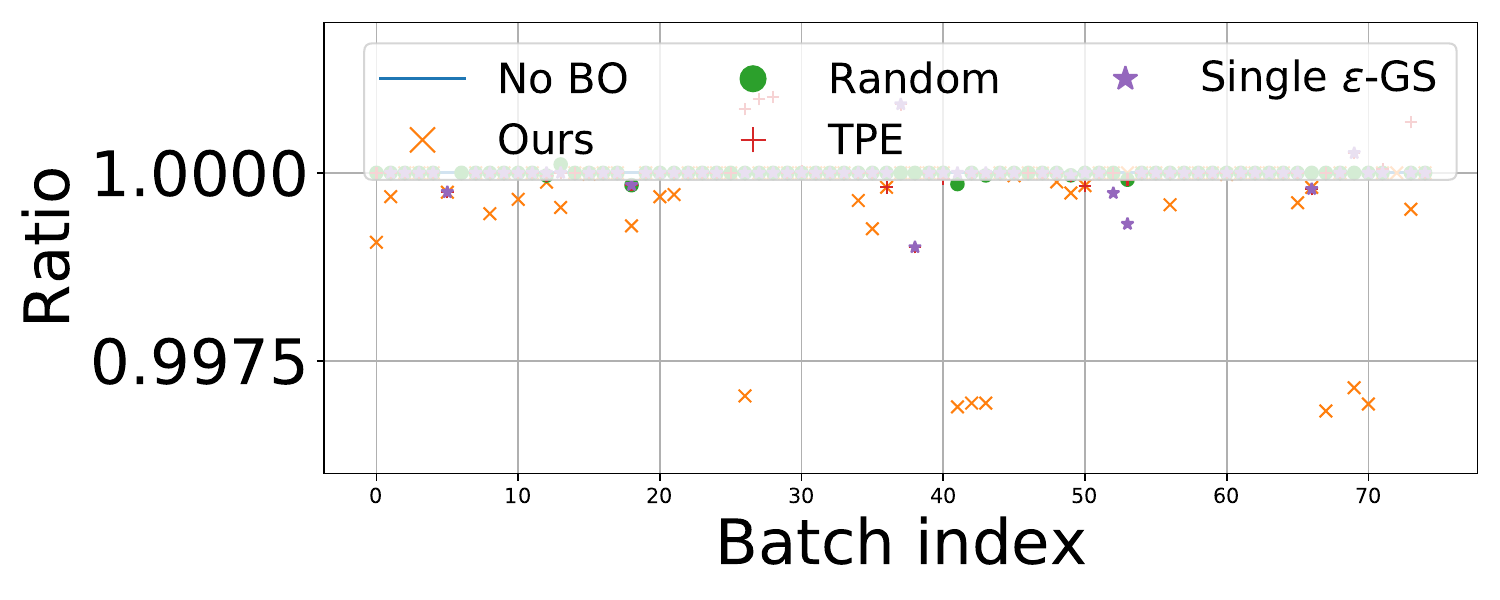}}
 \subfigure[GPT2 difference.]{
   \includegraphics[width=0.23\textwidth]{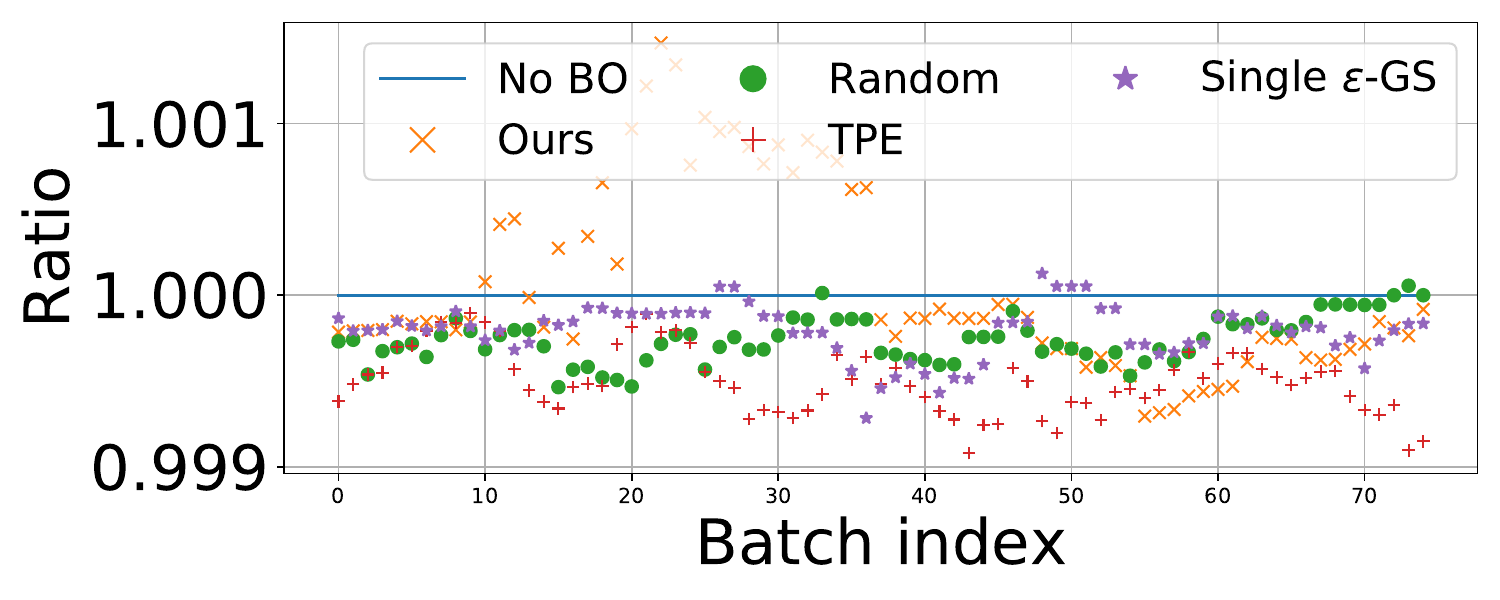}}
   \vspace{-3mm}
 \caption{Ratio of billed cost and expert prediction difference optimized by BO with different acquisition functions over no BO.
 }\label{fig_BO}
\end{figure}

\begin{figure}[!t]
 \centering
 \subfigure[Bert.]{
   \includegraphics[width=0.5\textwidth]{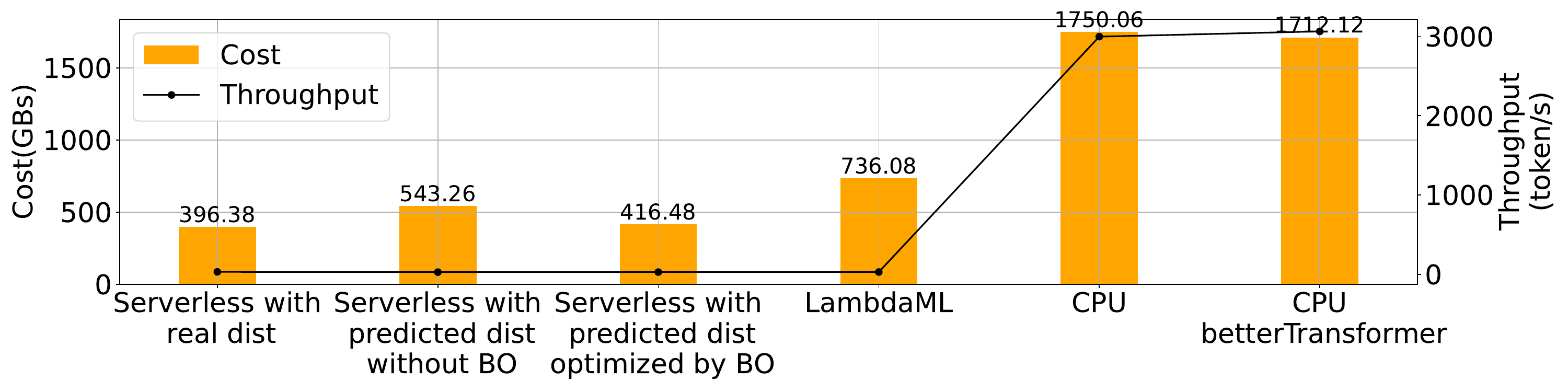}}
   \subfigure[GPT2.]{
   \includegraphics[width=0.5\textwidth]{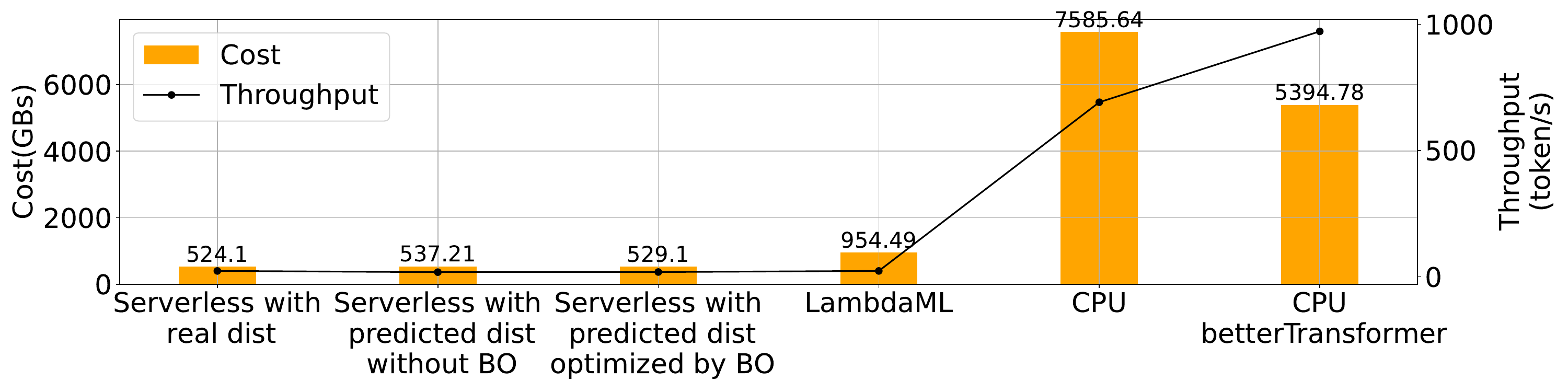}}
   \vspace{-3mm}
 \caption{Billed cost of all MoE layers and inverse of throughput under different expert selection distributions on AWS Lambda and CPU clusters. 
 }\label{fig_final_result}
\end{figure}

\subsection{Expert selection prediction} 

We first evaluate accuracy of expert selection prediction learned by our BO framework, by calculating the average absolute difference per expert between the real and predicted counts of tokens assigned to each expert.
Fig.~\ref{fig_expertselection} shows the 
difference across different MoE models, datasets, and tasks. For a task on a dataset, we use 95$\%$ of this dataset for profiling and evaluate the difference on 10,240 tokens in the dataset. Basic Bert MoE represents the Bert MoE model with 4 experts per MoE layer and top-1 routing for the fill-mask task on Enwiki8. The basic GPT2 MoE and the basic Bert2Bert MoE cases are similar. 
Other cases are variations based on these basic setups, e.g, GPT2 Lambda denotes the GPT2 basic setup but changes the dataset to Lambda \cite{lambda} dataset. 
Across all cases, our method outperforms expert prediction in Lina \cite{li2023accelerating}, as Lina only uses token ID as token feature while our method incorporates token ID, position ID, and attention ID to capture additional information 
for more accurate profiled probabilities. Compared to top-$1$ routing, the results of top-$2$ routing show that increasing the value of $k$ in top-$k$ gating significantly 
improves the prediction accuracy, as routing to more experts allows prediction mistakes in one expert to be corrected by the other. When the number of experts increases, the average prediction difference decreases. 

\subsection{Scatter-gather communication}
We next evaluate performance of our different scatter-gather communication designs. 
We allocate 3008MB of memory to each serverless function and use no expert replicas. 
Fig.~\ref{fig_comm} shows the billed cost of MoE layers and throughput of the entire MoE model 
under different communication methods. 
The results verify that the optimal scatter-gather communication method varies depending on the number of tokens. 
For 256 tokens, direct transfer performs best for both Bert MoE and GPT2 MoE. As the number of tokens increases, either pipelined or non-pipelined indirect transfers may perform better, while direct transfer becomes impractical due to payload size limitations; 
throughput increases because the time for model downloading and serverless function warm-up is distributed over more tokens. 

\subsection{ODS algorithm}
We 
deploy MoE model for
inference using 
10,240 tokens on AWS Lambda.
Fig.~\ref{fig_ODS} illustrates the billed cost of all MoE layers using our ODS algorithm, an MIQCP method and a random selection method, under different inference throughput targets. The MIQCP method uses one MIQCP solver to directly solve (\ref{formulation_v1}), and the random method randomly selects the communication method at each MoE layer. 
We set the target throughput by dividing 10240 tokens by the end-to-end latency limit specified in (\ref{formulation_v1}). 
The time limit for searching the optimal solution using the MIQCP approach is set to 180s; for the ODS algorithm with three MIQCP solvers, the search time limit is set to 60s. 
Our ODS algorithm outperforms other methods. 
At higher target throughputs, the MIQCP method fails to derive an optimal solution within 180s. 

\subsection{BO algorithm}
For BO learning, we use 10,240 tokens from the Enwiki8 dataset to simulate the inference requests, set $Q=1000$ key-value pairs to update in each BO iteration.
After BO learning, we test on 100 batches of inference requests from the dataset with batch size of 10240 tokens. Due to the long deployment time on AWS Lambda, we use simulation for this set of evaluation. 

Fig.~\ref{fig_BO} shows the ratio of the billed cost and expert prediction differences optimized by BO with different acquisition functions, to that of no BO, respectively. No BO means that we do not use the BO algorithm to adjust our expert predictor, and the expert selection is predicted by the unadjusted predictor. 
The random method randomly adjusts key-value pairs, the single $\epsilon$-greedy sampler uses the same $\epsilon$ for all dimensions of the variables, and the TPE \cite{bergstra2011algorithms} method samples on promising regions of variable range 
based on probabilistic modeling.
Our multi-dimension ${\bf \epsilon}$-GS performs best in terms of the billed cost 
for both Bert MoE and GPT2 MoE models. 
For BERT MoE, our method achieves the highest expert prediction accuracy. 
For GPT-2 MoE, our algorithm does not always yield the lowest prediction difference, 
possibly due to the increase of the number of tokens predicted for the most popular experts: our BO algorithm sets a larger $\epsilon$ for key-value pairs where memory is insufficient due to underestimation of expert popularity, resulting in an overestimation of the most popular experts that are most likely to exceed memory limits.

\subsection{Algorithm overhead}

We dissect the 
execution time of the expert selection predictor, the ODS algorithm and the BO algorithm. For expert selection predictor, the time of profiling 100 batches of data is around 28.89 seconds, and the prediction time on 10 batches is around 20.31 seconds. The execution time of the ODS algorithm with three MIQCP solvers is around 2.27s. Our BO algorithm requires around 62.15s per iteration and converges in around 1257.89s. 

\subsection{Overall performance}  
We 
deploy MoE models on AWS Lambda and a CPU cluster for inference serving of 10,240 tokens. The CPU cluster consists of two 64-core AMD EPYC CPUs with 512GB of DRAM. 

Fig.~\ref{fig_final_result} compares the billed cost of all MoE layers and inference throughput of the entire MoE model 
under different expert selection distributions (regarding the count of tokens assigned to each expert), 
among: 
(1) {\em Serverless with predicted distribution optimized by BO:} the optimal MoE deployement produced by our BO framework;
(2) {\em Serverless with real expert selection distribution:} the optimal MoE deployment produced based on ground truth of expert selections in the MoE inference; 
(3) {\em Serverless with predicted distribution without BO}: the optimal MoE deployment produced using predicted expert selections 
that is not adjusted by the BO algorithm;
 (4) {\em LamdaML \cite{lambdaml}}, which uses the maximum memory allocation for each serverless function on AWS Lambda (3008MB) for inference serving, requires no expert prediction, and uses no replicas for each expert;
 (5) {\em CPU:} the MoE model is deployed in the CPU cluster, 
 with all experts at each MoE layer executing concurrently, requiring no expert prediction;
 (6) {\em CPU betterTransformer:} MoE deployment in the CPU cluster accelerated by the CPU inference optimization method betterTransformer \cite{betterTransformer}, 
 through sparsity and fused kernels as Flash Attention. 

For both Bert MoE and GPT2 MoE, our serverless MoE inference design consistently results in lower billed costs, as compared to MoE inference on the CPU cluster. Specifically, serverless MoE inference with predicted expert selection reduces the billed cost by at least 75.67$\%$ compared to CPU cluster-based serving. The throughput in serverless-based MoE serving remains significantly above human reading levels of 3.3 tokens per second. 
The lower throughput in serverless MoE serving compared to CPU cluster-based serving is mainly because non-MoE layer computation is limited to the 3008MB memory size of each serverless function, 
which is far less than the 512GB available in a common CPU cluster. 
Among serverless options, the predicted expert selection distribution optimized by BO outperforms both non-BO methods and over-provisioning with LambdaML. The BO-optimized expert distribution not only reduces the billed cost by at least 43.41$\%$ compared to LambdaML with at most an 18.76$\%$ decrease in throughput, but also closely aligns with the cost of deployment using the real expert distribution.

\section{Conclusion}
This paper studies optimized MoE model deployment and distributed inference serving on a serverless platform, that effectively predicts expert selection, schedules communication with model execution, and minimizes the overall billed cost of serving MoE models. We propose a Bayesian optimization framework with multi-dimensional $\epsilon$-greedy search to learn expert selections and optimal MoE deployment. 
We design a novel Bayesian decision-making approach for expert selection prediction, 
propose several scatter-gather communication designs for serverless platforms, 
and design an ODS algorithm to decide optimal deployment of distributed MoE inference on a serverless platform. 
Experiments on AWS lambda validate our designs in reducing the billed cost of all MoE layers by at least 75.67$\%$ compared to CPU clusters while maintaining satisfactory inference throughput.  
As compared to LambdaML in serverless computing,
our designs achieve 43.41$\%$ lower cost with a throughput decrease of at most 18.76$\%$. 
\section*{Acknowledgment}

This work was supported in part by grants from Hong Kong RGC under the contracts C7004-22G (CRF), C6015-23G (CRF), 17204423 (GRF), 16210822 (GRF) and T43-513/23-N (TRS).

\bibliographystyle{IEEEtran}
\bibliography{refer}

\end{document}